\documentclass[10pt,twocolumn]{article} 

\usepackage{simpleConference}
\usepackage{times}
\usepackage{graphicx}
\usepackage{amssymb}
\usepackage{url,hyperref}
\usepackage{comment}
\usepackage{float}
\usepackage[export]{adjustbox}
\usepackage{setspace}

\usepackage{caption}
\renewcommand{\baselinestretch}{1.5}

\begin{document}

\title{Eight exoplanet candidates in SAO survey}
\author{O.~Ya.~Yakovlev$^{*1,2}$, A.~F.~Valeev$^{2,3,4}$, G.~G.~Valyavin$^{2}$, A.~V.~Tavrov$^{1,5}$,\\
		V.~N.~Aitov$^{2}$,  G.~Sh.~Mitiani$^{2}$, G.~M.~Beskin$^{2,6}$, V.~V.~Vlasyuk$^{2}$,\\ O.~I.~Korablev$^{1}$, G.~A.~Galazutdinov$^{4,2}$, E.~V.~Emelyanov$^{2}$, T.~A.~Fatkhullin$^{2}$,\\
		V.~V.~Sasyuk$^{6}$, A.~V.~Perkov$^{7}$, S.~F.~Bondar$^{\dag}$$^{7}$, 
		T.~E.~Burlakova$^{2,4}$, S.~N.~Fabrika$^{2}$, I.~I.~Romanyuk$^{2}$\\
\\
$^{1}$Space Research Institute, Moscow, 
$^{2}$Special Astrophysical Observatory, Nizhnii Arkhyz,\\ 
$^{3}$St.~Petersburg University, St.~Petersburg,
$^{4}$Crimean Astrophysical Observatory, Nauchny,\\
$^{5}$Moscow Institute of Physics and Technology, Dolgoprudny,
$^{6}$Kazan Federal University, Kazan,\\
$^{7}$Research and Production Corporation ``Precision Systems and Instruments'', Moscow\\
$^{*}$yko-v@ya.ru\\
\\Accepted by Astrophysical Bulletin, November 29, 2022
}

\twocolumn[
\maketitle
\begin{abstract}
	Here we present eight new candidates for exoplanets detected by the transit method at the Special Astrophysical Observatory of the Russian Academy of Sciences. Photometric observations were performed with a \mbox{50-cm} robotic telescope during the second half of 2020. We detected transits with depths of \mbox{$\Delta m = 0\,.\!\!^{\rm m}056$--$0\,.\!\!^{\rm m}173$} and periods $P = 18\,.\!\!^{\rm h}8$--$8\,.\!\!^{\rm d}3$ in the light curves of stars with magnitudes of \mbox{$m=14\,.\!\!^{\rm m}3$--$18\,.\!\!^{\rm m}8$}. All considered stars are classified as dwarfs with radii of \mbox{$R_* =0.4$--$0.6 R_{\odot}$} (with the uncertainty for one star up to $1.1\,R_{\odot}$). We estimated the candidate radii (all are greater than 1.4 times the Jovian radius), semi-major axes of their orbits (0.012--0.035~AU), and other orbital parameters by modelling. We report the light curves with transits for two stars obtained in 2022 based on individual observations.
\end{abstract}
]

\section{Introduction}
The first optical robotic telescope with 50-cm mirror at the Special
Astrophysical Observatory of the Russian Academy of Sciences
(SAO RAS) aimed at studies of exoplanets started operating in 2020.
Its primary purpose is to find new exoplanet candidates.  Transit
photometry technique is used~--- an exoplanet is found and further studied 
by analyzing photometric data acquired during the transit.
It also observes stars with known confirmed exoplanets.

To discover exoplanet candidates, photometric monitoring of a selected
sky area was performed during the second half of 2020 (hereafter referred to as survey-mode observations).
So far, no confirmed exoplanet of exoplanet candidate has been found in this field according to the NASA Exoplanet
Archive data \cite{NASA}.
As a result of the analysis of about  30\,000 light curves of
\mbox{$m=13\,.\!\!^{\rm m}5$--$19\,.\!\!^{\rm m}5$} stars in the
area studied we found eight objects exhibiting periodic brightness
variations typical of planet transit events (in terms of
periodicity, amplitude, duration, and form).
The objects  are designated as SOI-N, i.e ``SAO Object of
Interest'' followed by the candidate number $N$. Since the beginning
of 2022 dedicated observations of these objects are performed at
the SAO RAS including multicolor photometry (50-cm and
\mbox{1-m telescopes}) and spectroscopy (the 6-m telescope BTA).

The area that we selected for preliminary study is centered on the magnetic white dwarf \mbox{WD\,0009$+$501} \mbox{($m_G\approx14\,.\!\!^{\rm m}2$)} with eight-hour periodic light  variations with an amplitude of $\Delta m\approx0\,.\!\!^{\rm m}01$ \cite{WD}. This result is confirmed by the structure of the light curve obtained. In the process of our search for transit events we detected more than 100 new variable stars with $\Delta  m\approx0\,.\!\!^{\rm m}01$--$0\,.\!\!^{\rm m}1$ light variations. We will present results of their studies in a separate paper.

The transits of the exoplanet candidates discussed in this paper have
depths in the range \mbox{$0\,.\!\!^{\rm m}06$--$0\,.\!\!^{\rm m}17$}.
Such a brightness decrease corresponds to a transit of a giant planet
the size of 1.5--2 times that of Jupiter across the disk of a
solar-radius or smaller star \cite{Frontier}.

The duration of survey-mode observations (5~months) and their duty cycle (on the average, every third night) determine the possible intervals of detectable transit-event periods in the objects found~--- 1--2~days and up to slightly more than a week. Such a period corresponds to planets with orbit semi-major axes smaller than 0.1~AU. These observed parameters (the depth and period of brightness decrease) are consistent with earlier preliminary estimates \cite{WD} and with statistical data for confirmed exoplanets discovered using the same technique (i.e., via transit observations on ground-based instruments) taken from the Exoplanet Archive data \cite{NASA} and \cite{Frontier}.

For each of the exoplanet candidates presented in this paper light curves covering three or more complete transits are available that were fitted to model curves in order to determine the radius of the candidate and its orbital parameters, primarily the semi-major axis of the orbit. According to these models, the observed transit events can be caused by transits of planets across the stellar disk. The currently available data are insufficient to confidently reject the hypothesis about the planetary nature of the candidates considered.

Here we publish coordinates and other information about stars SOI-1--SOI-8, as well as the ephemerides  of the transit events. See \cite{Frontier}
for details of observations and data reduction procedure.
Further observations with other instruments will be useful for confirming or discarding the hypothesis about the planetary nature
of the candidates discovered.

\section{Observations and data processing}

Observations were carried out in two stages. At the first stage, from
August 25, 2020 to January 21, 2021, light curves for more than  35\,000 stars
were acquired with a total duration of up to  $22\,.\!\!^{\rm d}8$ over 56 nights.
Observations were performed for a total of 84 nights.
These were wide-field (\mbox{$\Delta\alpha=2\,.\!\!^\circ45$},
\mbox{$\Delta\delta=1\,.\!\!^\circ56$}) survey-mode observations on
a large scale ($1\,.\!\!^{\prime\prime}34/$pixel) performed on a 50-cm telescope
with 60-s exposure using a detector located in the primary focus. Since
February 2022 the second stage began~--- $BVRI$-band observations of individual
objects discovered during the survey with same and another 50-cm telescope
(equipped with a detector placed at the secondary focus with a scale of
$0\,.\!\!^{\prime\prime}93/$pixel after binning $2\times2$) (see Table~\ref{table:journal}).
We will use the results of more precise transit photometry based on
the data from the 1-m telescope and the SOI spectra acquired with the 6-m telescope
to refine the parametes of exoplanet candidates and publish them in the future.

\begin{table}[H]
	\caption {Log of individual observations\\of exoplanet candidates}
	\centering 
	\begin{tabular}{c|c|c}
		\hline
		Date 			& Filters 		& Object  \\
		\hline
		22.02.11    	& B, V, R 		& SOI-3\\
		\hline
		22.02.13    	& B, V, R 		& SOI-3\\
		\hline
		22.05.29 		& B, V, R, I 	& SOI-8\\
		\hline
		22.06.02 		& B, V, R, I 	& SOI-3\\
		\hline
		22.06.05 		& V, I 			& SOI-8\\
		\hline
		22.06.28 		& V, R, I 		& SOI-2*\\
		\hline
		22.07.04 		& R	 			& SOI-7, SOI-8 \\
		\hline
		22.07.08 		& I, V 			& SOI-7, SOI-8\\
		\hline
	\end{tabular}\\
	\label{table:journal}
	* transit was not observed
\end{table}

Photometric data were processed in pipeline mode using standard procedures: dark-frame subtraction (no flat field frames
were obtained), identification of sources in the frame and coordinate fitting, aperture photometry, and cross-matching.
To perform data reduction we used a  {\tt Python} script with standard programs: {\tt Astrometry.net} \cite{Astrometry},
{\tt SExtractor} \cite{SExtractor},
and {\tt  CCDPACK} \cite{CCD}.
	
These procedures were followed by light-curve calibration. We used bright stars ($14\,.\!\!^{\rm m}0$--$15\,.\!\!^{\rm m}5$) with high correlation coefficient between instrumental light curves and the star compared, and with many observational data points (close
to the maximum number).

\section{Analysis of light curves}

\begin{figure}[H]
	\centering
	\includegraphics[scale=0.45]{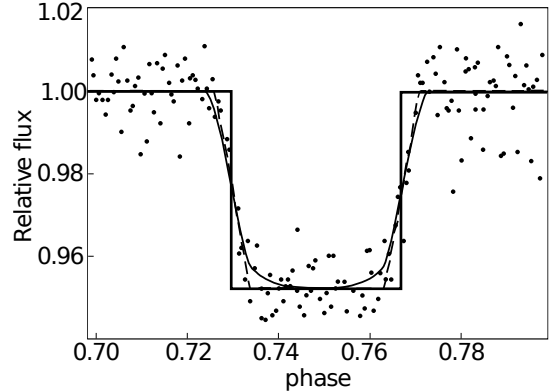}
	\caption{Models fitting the transit events: box-shape \cite{BLS}, \cite{astropy} (the thick line), trapezium-shaped (the dashed line) and {\tt batman} \cite{batman} (the thin line). The dots show a fragment of the SOI-3 light curve with a transit.
	}
	\label{fig:models}
\end{figure}

We investigated the resulting light curves for periodic brightness dips expected in the case of an exoplanet transit across the
stellar disk. To this end, we employed the simple Box Least Squares algorithm (BLS) \cite{BLS}, which we used to select the
parameters of the two-level model fitting the light curve phase-folded with the trial period \cite{Frontier}.
In this model the outside-transit light curve is approximated by a single (zero) level, and the inside-transit part of the light curve, by another level (Fig.~\ref{fig:models}). \linebreak

The search for light curves with transit events is performed by analyzing the periodograms that statistically characterize the
agreement between the phased light curve and the adopted model. The main parameter determined at this stage using the BLS algorithm
is the period $P$, i.e., the time interval between two consecutive transits (the orbital period of the body that causes the
darkening). The remaining parameters of this model (duration $\tau$, depth $\Delta m$, and the position of the central transit
point $t_0$) are refined using other models described below.

We use {\tt astropy} ({\tt Python}) library \cite{astropy}
to implement the BLS algorithm. The input parameters are the light
curve and the period $P$ and transit duration $\tau$ grids. We proceeded from the duration and duty cycle of observations, photometric precision, statistical data about the parameters of known exoplanets, and simple estimates  (see Fig.~9 in \cite{Frontier}) to select the following intervals of inferred parameters: $P=0.5$--$5$~days and \mbox{$\tau=0.1$--$5$~hours}
(candidates SOI-4 and SOI-5 with periods longer than five days were initially found with shorter multiple periods).

Hereafter to determine the parameters of transit events and the physical and orbital parameters of exoplanet candidates we
convert the relative magnitudes $\Delta m$ into relative flux units $\Delta F$. The zero (outside-transit) level is set equal to~1.

The next approximation is the trapezium-shaped light curve. It differs from the previous model (BLS) by the presence of
transition fragments between the zero level and the  ``flat bottom'', during which only a part of the body lies in front of
the stellar disk. This model is described by four time parameters:
the beginning and end of the transit, $t_1$, $t_4$, and the beginning and end of maximum brightness decrease phase, $t_2$, $t_3$.
The depth of the transit  $\Delta F$ is, like in the previous model, determined as the median value in the  $t_2$--$t_3$ interval.

The radius of the body, $R_{{p}}$, and the semi-major axis of its axis, $sma$, in the units of stellar radius $R_*$
can be determined from the shape and amplitude of the light curve
\mbox{($t_{1}$, $t_2$, $t_3$, $t_4$, $\Delta F$)}
assuming that the orbit is circular \cite{Seager}:
\begin{equation}
	\label{eq:one}
	R_{{p}}/R_* = \sqrt{\Delta F},
\end{equation}
\begin{equation}
	\label{eq:two}
	sma/R_* = \sqrt{\frac{(1 + \sqrt{\Delta F})^2 - b^2 (1 - f_1)}{f_2}},
\end{equation}
where $f_1 = \sin^2(\pi\,t_T/P)$ and \mbox{$f_2 = \sin^2(\pi\, t_F/P)$} are functions that depend on the total
duration of the transit, $t_T = t_4 - t_1$, and the duration of the maximum transit depth phase, $t_F = t_3 - t_2$,  and
\linebreak \mbox{ $b = f(\Delta F, f_1, f_2)$} is the projected distance between the planet and the center of the star at time $t_0$
(see equations~(6)--(8) in \cite{Seager}:

The third model, which describes best the transit compared to the above two models is  {\tt batman} (BAsic Transit Model cAlculatioN)
\cite{batman}.
The light curve is determined by eight parameters: the central point of the transit, $t_0$, radius of the planet
$R_{{p}}/R_*$ and semi-major axis of its orbit $sma/R_*$ in the units of stellar radius; inclination $i$ (angle between the
orbital and sky planes); eccentricity $e$ and longitude of the ascending node $\Omega$ of the orbit, and the coefficients of
the quadratic limb-darkening function of the star, $a$ and $b$.

We used the algorithm of the first  (Box-shape) model to find more than 30 stars with transit events. For each of these stars we
determined the transit epoch ${\mathrm{JD}_0}$, period $P$, and the depth $\Delta m$ and duration  $\tau$ of the brightness dip.
We then used the inferred estimates of these parameters to distinguish stars with exoplanet candidates (SOI) from eclipsing variables
with $\Delta m \geq 0\,.\!\!^{\rm m}2$, $P \geq 5^{\rm h}$, $\tau/P \geq 0.25$ and  \mbox{V-shaped} transit light curves or with different
depths  $\tau$ of odd and even transits.

We determined the parameters of the second (trapezium-shaped) and third  ({\tt batman}) models via the maximum-likelihood method.
We used the   \mbox{$t_{1}$--$t_{4}$} and
$\Delta F$ estimates of the second model to determine the intervals of inout parameter values for the third model:
($t_0$, $R_{{p}}/R_*$ from Eq.~(1) and $sma/R_*$ from Eq.~(2)).
The intervals of orbital parameters are   $i=75$--$90^\circ$, \mbox{$e =0$--$1$} and \mbox{$\Omega=0$--$360^\circ$}.
We adopted the coefficients $a$ and $b$ of the star's quadratic limb-darkening function from the table published by \cite{Claret}
based on the following stellar parameters:
effective temperature  $T_{\mathrm{eff}}$, surface gravity $g$,
and metallicity [M/H]  \cite{GAIA1}, \cite{GAIA2}.

The aim of this modeling stage was to estimate the period, duration, and depth of the transit, the radii of exoplanet candidates, and
the semi-major axes of their orbits in order to test the hypothesis of the planetary nature of these bodies (other, orbital, parameters
are accompanying quantities). We will perform further refining modeling with error estimates based on photometric and spectroscopic data
obtained with the 1-m \mbox{Zeiss-1000} and 6-m telescopes of the SAO RAS.

\section{Results}

\begin{figure*}
	%\vspace{-1mm}
	%\medskip
	\caption{
		Light curves of eight exoplanet candidates during the entire time of the monitoring. The solid horizontal lines show the zero level and maximum transit depth, and the dashed lines,
		the $\pm3\sigma$ levels. The labels under the light curves indicate the nights during which the transit was observed:
		only the beginning (leftward arrow), only the end (rightward arrow), mid-transit with the beginning or end  (``plus'' symbol)	or complete transit (upward arrow).
	}
	\includegraphics[width=1\textwidth,]{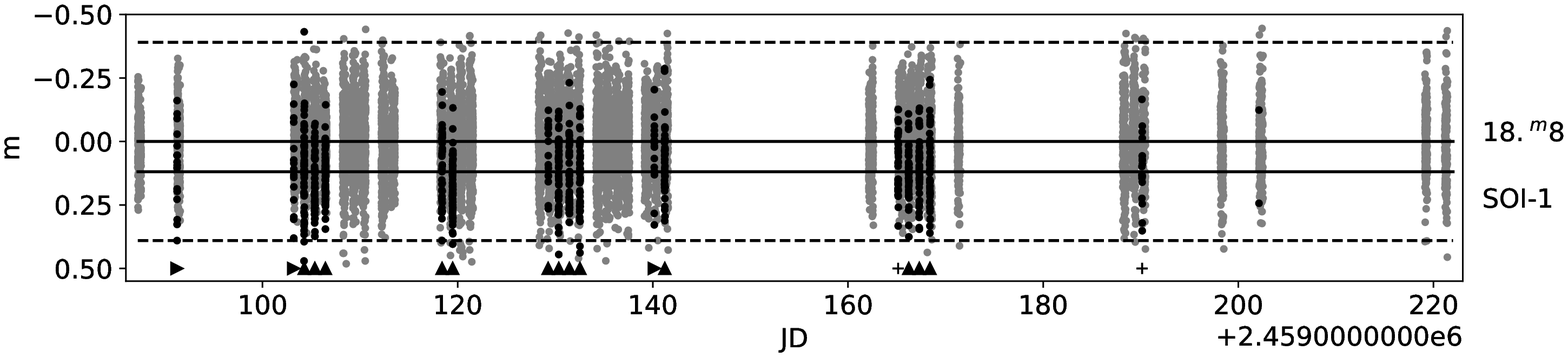}
	\includegraphics[width=1\textwidth,]{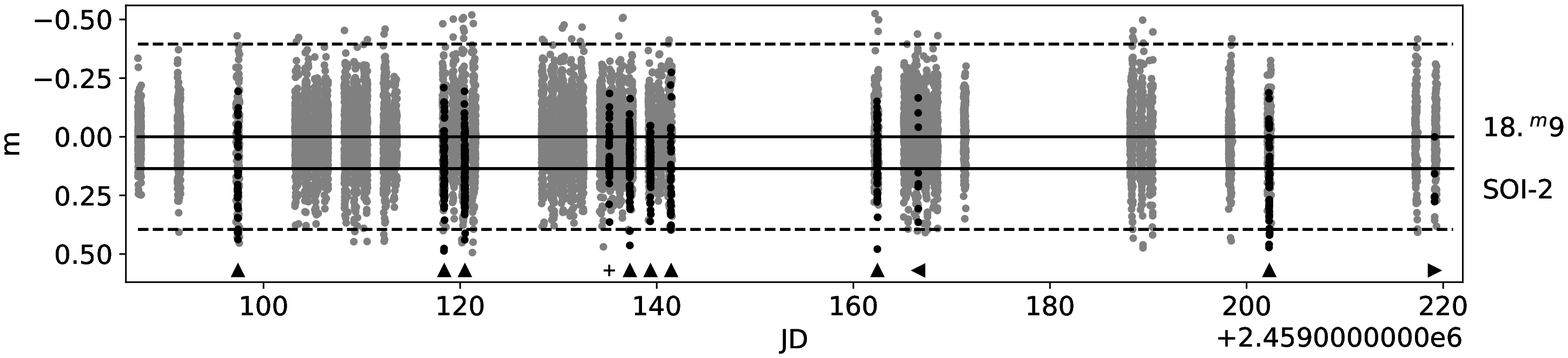}
	\includegraphics[width=1\textwidth,]{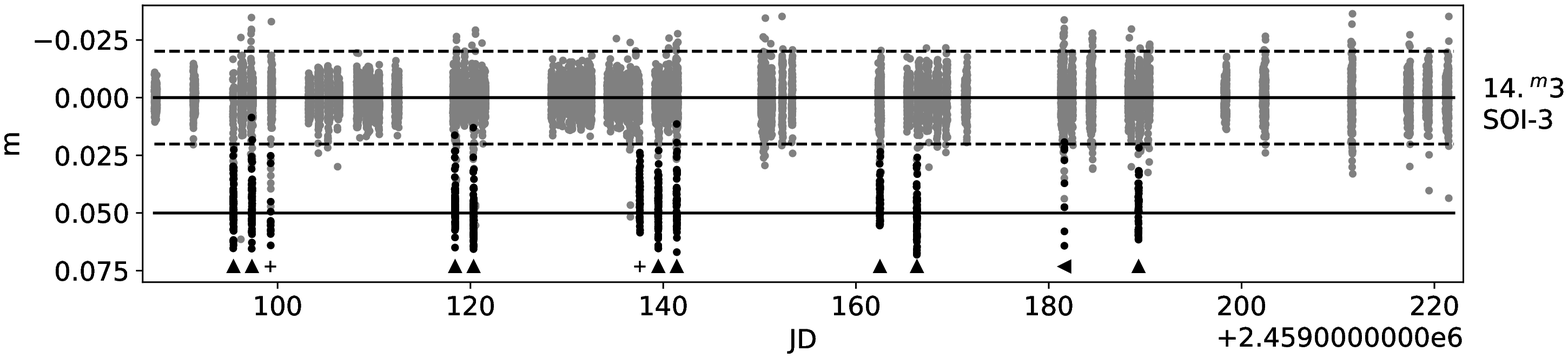}
	\includegraphics[width=1\textwidth,]{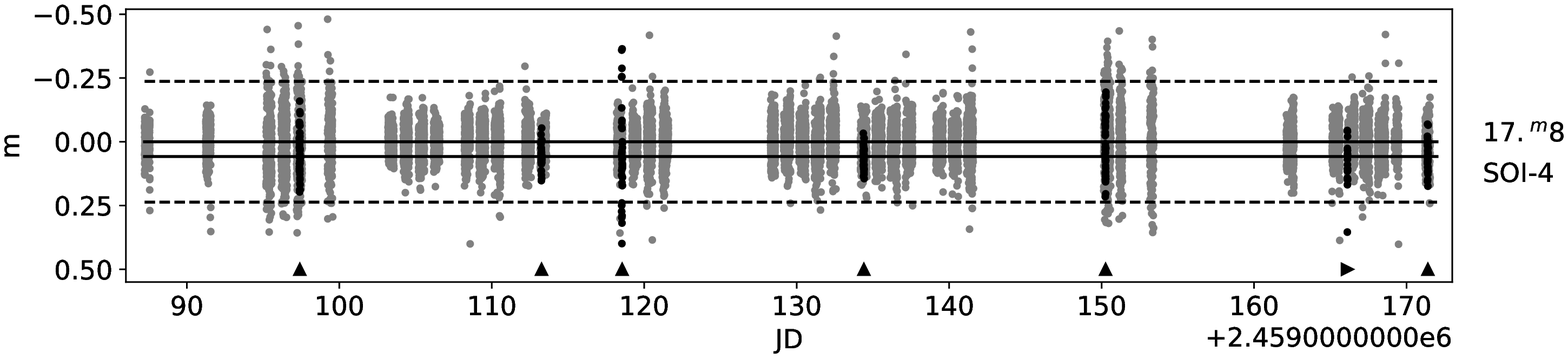}
	\label{fig:LC_all}
\end{figure*}

Analyzing the light curves of stars in the field studied, we found eight stars exhibiting periodic light dips characteristic of transit events. Fig.~\ref{fig:LC_all}
shows the light curves of these stars (with individual nightly trend subtracted and outliers and data obtained during nights with bad weather conditions removed).
The transit depth  $\Delta m$ for the brightest stars \mbox{SOI-3}, SOI-5 ($m<15^{\rm m}$), and \mbox{SOI-8} ($m\approx17^{\rm m}$)
exceeds $3\sigma$.
The other candidates are fainter ($m\approx18^{\rm m}$--$19^{\rm m}$), we estimate the transit depths for them as $\Delta m < 3\sigma$.

We present the data for stars SOI-1--SOI-8 including IDs from
Gaia\,DR3 (Vallenari et al. 2022), magnitudes, coordinates, physical
parameters and the inferred estimates of the observed parameters of
transit events  (zero epoch, period, and depth) below in this Section
and also in summary Tab.~\ref{table:params_main}, \ref{table:params_phys}. Partial differences of the values of the above parameters for SOI-1--SOI-5 from the estimates earlier published by  \cite{Frontier} can be explained by  the use of a different Gaia data release (DR3 instead of EDR\,3) and different transit models~--- the trapezium-shaped and {\tt batman} model \cite{batman}.

\begin{table*}
	\caption {
		Parameters of stars with exoplanet candidates:
		Gaia~DR3~ID, coordinates RA and Dec, and apparent magnitude
		$m_G$ from Gaia database \cite{GAIA2}
		and parameters of transit events, including
		zero epoch ${\mathrm{JD}_0}$ of one of the transits, period $P$
		between the transits observed, duration $\tau$ and maximum depth
		$\Delta m$ of the transit}
	\begin{tabular}{c|c|c|c|c|c|c|c|c}
		\hline
		SOI             & Gaia\,ID       &  RA, hh:mm:ss    & Dec, dd:mm:ss         & $m_G$, mag     & ${\mathrm{JD}_0}$     & $P$, h       & $\tau,$ h        & $\Delta m,$ mag  \\
		\hline
		1   & 395249244499720320        & 00:10:30.08     & +50:27:54.0      & 18.81     & 2459202.090069        &  26.085        & 1.56      & 0.14 \\
		2   & 395239868590825472        & 00:12:31.18     & +50:27:21.3      & 18.83     & 2459219.105861        &  50.358        & 2.97      & 0.17 \\
		3   & 395281753112434304        & 00:08:26.77     & +50:28:23.9      & 14.30     & 2459189.314361        &  45.988        & 2.17      & 0.06 \\
		4   & 395220833295162112        & 00:11:10.60     & +50:16:21.6      & 17.81     & 2459171.403981        & 126.844        & 1.49      & 0.08 \\
		5   & 395229079635160320        & 00:13:17.88     & +50:15:43.9      & 15.17     & 2459221.434447        & 198.309        & 3.81      & 0.06 \\
		6   & 395282813962513536        & 00:07:59.73     & +50:31:19.9      & 17.96     & 2459202.386315        &  49.645        & 2.79      & 0.06 \\
		7   & 394596585565904000        & 00:16:33.19     & +50:41:10.1      & 18.24     & 2459236.309815        &  97.670        & 2.55      & 0.15 \\
		8   & 394573186584164096        & 00:16:01.05     & +50:32:24.1      & 16.94     & 2459221.159153        &  18.768        & 1.47      & 0.09 \\
		\hline
	\end{tabular}
	\label{table:params_main}
\end{table*}
\renewcommand{\baselinestretch}{5}

\subsection*{SOI-1}
The candidate SOI-1 with $m_G=18\,.\!\!^{\rm m}81$
was found to exhibit brightness dips with the period of \mbox{$P = 26\,.\!\!^{\rm h}09$},
duration \mbox{$\tau = 1\,.\!\!^{\rm h}56$}, and depth \mbox{$\Delta m=0\,.\!\!^{\rm m}144$} (Fig.~\ref{fig:SOI_1}).
A total of 12 complete transits (from the beginning to the end) were detected for this star during survey-mode observations.

The limb-darkening coefficients for \mbox{SOI-1} are:
$a~=~0.354$ and $b~=~0.374$ (with $\log g~=~4.90$ and \linebreak \mbox{$\log {\rm (M/H)}=-1.69$)}. The fitted model transit
curve corresponds to the following parameters:
\mbox{$\phi_0\!=\!0.7501$}, $R_{{p}}/R_*=0.38$, $sma/R_*=7.85$, \mbox{$i=83\,.\!\!^\circ3$,} $e=0.60$, \linebreak and $\Omega= 307^\circ$.

\subsection*{SOI-2}
The period of transits for  SOI-2 with \mbox{$m_G=18\,.\!\!^{\rm m}83$} was initially estimated as
$P=25\,.\!\!^{\rm h}2$ (this period is reported in \cite{Frontier}).
After the subtraction of the daily trend the period corresponding to the
maximum of the periodogram increase twofold (Fig.~\ref{fig:SOI_2_PG}).

The phased light curve with \mbox{$P=50\,.\!\!^{\rm h}36$} reveals a secondary minimum (Fig.~\ref{fig:SOI_2},
$\phi\approx 0.25$),
which was mistaken for the primary minimum along with the minimum at \mbox{$\phi\approx 0.75$}.
The transit duration and depth are equal to $\tau=2\,.\!\!^{\rm h}97$ and $\Delta m = 0\,.\!\!^{\rm m}173$, respectively. A total of eight complete transits
were registered during survey-mode observations.

The limb-darkening coefficients for  SOI-2 are derived to be \mbox{$a=0.22$}, $b=0.44$ ($\log g=4.77$ and \mbox{$\log {\rm (M/H)}=-2.52$)}.
The fitted model transit curve corresponds to  the following parameters: $\phi_0\!=\!0.7500$, $R_{{p}}/R_*=0.42$, $sma/R_*=8.47$,
$i=83\,.\!\!^\circ7$, $e=0.50$, and $\Omega= 219^\circ$.

 \subsection*{SOI-3}
SOI-3  is the brightest among the eight studied ($m_G=14\,.\!\!^{\rm m}30$). Its light
curve was found to exhibit brightness dips with the period, duration, and depth equal to  %\linebreak
$P = 45\,.\!\!^{\rm h}99$, $\tau = 2\,.\!\!^{\rm h}17$,
and \mbox{$\Delta m=0\,.\!\!^{\rm m}056$}, respectively (Fig.~\ref{fig:SOI_3}).
A total of eight complete and four partial transits were
detected during survey-mode observations.

Temporary flux increase by \mbox{$0\,.\!\!^{\rm m}01$--$0\,.\!\!^{\rm m}02$}
is observed at the periods of 16, 28, and 53  during the transit. This
effect can be due to the flare activity in the stellar atmosphere.

The limb-darkening coefficients for \mbox{SOI-3} are: \linebreak \mbox{$a=0.59$}, $b=0.19$ ($\log g\!=\!4.20$,
\mbox{$\log {\rm (M/H)}\!=\!-0.99$}). The fitted model transit curve corresponds to the following parameters:
$\phi_0=0.7489$, $R_{{p}}/R_*=0.24$, $sma/R_*=4.50$, $i=81\,.\!\!^\circ2$, $e=0.45$, and $\Omega= 90^\circ$.

In the case of  SOI-3 the light curves for three nights in 2022 were obtained with predicted transit event (Fig.~\ref{fig:SOI_3_post}).
All three detected transits were observed at phases  \mbox{$\phi<0.5$}, i.e., so far no predicted transit could  be observed completely. The beginning of the transit is present in the light curve.

\subsection*{SOI-4}
The star SOI-4 with $m_G=17\,.\!\!^{\rm m}81$ was found to exhibit transits with the period
%\linebreak
$P = 126\,.\!\!^{\rm h}85\approx5^{\rm d}6^{\rm h}$, duration \mbox{$\tau = 1\,.\!\!^{\rm h}95$},
and depth $\Delta m=0\,.\!\!^{\rm m}071$ (Fig.~\ref{fig:SOI_4}).
A total of six transits were detected during survey-mode observations.

The limb-darkening coefficients we derived for \mbox{SOI-4} are:
\mbox{$a=0.27$}, $b=0.48$ (with $\log g=4.63$,
\mbox{$\log {\rm (M/H)}\!=\!-0.33$}). The fitted transit curve corresponds to the following parameters:
$\phi_0=0.7493$, $R_{{p}}/R_*=0.30$, $sma/R_*=13.22$, $i=79\,.\!\!^\circ8$, $e=0.85$, and $\Omega= 195^\circ$.

\subsection*{SOI-5}
The star SOI-5 (ID\,395229079635160320 in Gaia\,DR3, ID\,395229079633073024 in EDR3)
with \mbox{$m_G=15\,.\!\!^{\rm m}17$} was found to manifest
transit events with the longest~--- more than one
\mbox{week~--- period} among the eight stars considered,
%\linebreak
\mbox{$P\! =\! 198\,.\!\!^{\rm h}31\!\approx\!8^{\rm d}6^{\rm h}$},
with the duration $\tau = 3\,.\!\!^{\rm h}81$ and depth \mbox{$\Delta m=0\,.\!\!^{\rm m}060$}
(Fig.~\ref{fig:SOI_5}).
One complete transit and three partial transits with phases up to
$\phi\approx 0.85$ were found during survey-mode observations.

The fitted model transit curve corresponds to the following parameters:
\mbox{$\phi_0=0.7501$}, $R_{{p}}/R_*=0.24$, $sma/R_*=14.34$, $i=85\,.\!\!^\circ4$, $e=0.30$, $\Omega= 170^\circ$, \linebreak  $a=0.25$, and $b=0.42$.
The parameters of the star  SOI-5 are unknown and that is why the parameters $a$ and $b$ were inferred in the process of
modeling.

\subsection*{SOI-6}
Transits events with the period $P = 49\,.\!\!^{\rm h}65$, duration
\mbox{$\tau = 2\,.\!\!^{\rm h}79$}, and depth $\Delta m=0\,.\!\!^{\rm m}060$ (Fig.~\ref{fig:SOI_6})
were found for the star SOI-6  ($m_G=17\,.\!\!^{\rm m}96$). During survey-mode observations
a total of four complete transits were registered.

The limb-darkening coefficients for  %\linebreak \mbox{SOI-6}
the object are: $a=0.25$, $b=0.42$ ($\log g\!=\!4.72$,
%\linebreak
\mbox{$\log{\rm (M/H)}\!=\!-2.78$)}. The fitted model transit curve corresponds to the following parameters:
$\phi_0=0.7413$, $R_{{p}}/R_*=0.25$, $sma/R_*=9.58$, $i=87\,.\!\!^\circ6$, $e=0.60$, and $\Omega= 238^\circ$.

\subsection*{SOI-7}
In studies of the star  SOI-7 with $m_G=18\,.\!\!^{\rm m}24$
transit events with the period $P = 97\,.\!\!^{\rm h}67\approx4^{\rm d}17^{\rm h}$,
duration \mbox{$\tau = 2\,.\!\!^{\rm h}55$}, and depth $\Delta m=0\,.\!\!^{\rm m}153$
were revealed (Fig.~\ref{fig:SOI_7}). During survey-mode observations
we registered a total of three complete transits
and one with transit phase before  $\phi\approx 0.5$.

The limb-darkening coefficients for \mbox{SOI-7} are estimated as $a=0.37$, $b=0.37$
($\log g=4.79$ and   \linebreak  \mbox{$\log {\rm (M/H)}=-1.52$}).
The fitted model transit curve corresponds to the following parameters:
$\phi_0=0.7493$, $R_{{p}}/R_*=0.39$, $sma/R_*=8.24$, $i=81\,.\!\!^\circ5$, $e=0.55$, and $\Omega= 40^\circ$.

\subsection*{SOI-8}
The star SOI-8 with $m_G=16\,.\!\!^{\rm m}94$
was found to exhibit transit events with the shortest period~--- less than one day or
$P = 18\,.\!\!^{\rm h}77$~--- among the eight stars studied (Fig.~\ref{fig:SOI_8}).
The transit duration and depth
are \mbox{$\tau = 1\,.\!\!^{\rm h}47$} and $\Delta m=0\,.\!\!^{\rm m}089$, respectively.
A total of 19 complete and three partial transits were detected
during survey-mode observations.

The limb-darkening coefficients for \mbox{SOI-8} are: $a=0.50$, $b=0.25$ ($\log g\!=\!4.72$,
\mbox{$\log {\rm (M/H)}\!=\!-1.03$)}.
The fitted model transit curve corresponds to the following parameters:
$\phi_0=0.7452$, $R_{{p}}/R_*=0.30$, $sma/R_*=4.59$, $i=82\,.\!\!^\circ8$, $e=0.05$, and $\Omega= 273^\circ$.

In 2022 light curves for \mbox{SOI-8} were obtained for four nights during which transit events were predicted
(Fig.~\ref{fig:SOI_8_post}).
Two of these were observed completely in two filters \mbox{($V$, $R$)}, and two,
partially in three filters ($B$, $V$, $R$).

Besides transits, SOI-8 also manifests brightness variations (inferred from an analysis of the Lomb--Scargle periodogram in \cite{LombScargle})
with the period \mbox{$P=18\,.\!\!^{\rm h}688$} and amplitude $\Delta m \approx 0.1$ (Fig.~\ref{fig:SOI_8_PG} and Fig.~\ref{fig:SOI_8_Var}).
This period is shorter than the period of transits \mbox{$P_{\mathrm{tr}}=18\,.\!\!^{\rm h}768$} by about  4.5~minutes.
Fig.~\ref{fig:SOI_8_Var} also shows the phase-folded light curve for  $P=P_{\mathrm{tr}}$ with the transit-event phase excluded.

\section{Discussion}
In  the cases where two transits of the same shape were registered with consistent periodicity over a long time interval
the hypotheses that light variations may be caused by star spots, phenomena in terrestrial atmosphere, or arose in the
process of observations or data reduction can be discarded. Hence the brightness dip is cased by the apparent transit of a
body that is gravitationally bound to the star and may be a star, brown dwarf, or planet. These conditions are satisfied for
all SOI: four or more transits (at least three complete ones) were detected over several months, for two objects
(SOI-3 and SOI-8) including transits observed in 2022 at times predicted based on the data of 2020 survey-mode observations, and
recurrent shape of transits (during transit time the light curve deviates from the mode curve by no more than $3\sigma$).
A transit is observed at any time epoch separated from the zero epoch by integer number of periods (in the cases where
data are available).

The transit technique makes it possible to determine only the radius, whereas the mass of the body remains unknown. If the transit is caused by a body comparable in size with gas giants, we can exclude the presence of a stellar component. If the radius of the body is significantly smaller than the radius of Jupiter, we can conclude that the body
is an exoplanet. All candidates found in this study have radii greater than that of Jupiter and therefore we cannot rule out the possibility that they may be brown dwarfs or small stars.

Mass  is the physical parameter that distinguishes planets from brown dwarfs.
The adopted conventional boundary is set at $13\,M_{\mathrm{Jup}}$: if the mass of the object exceeds this value, the hypothesis of its planetary nature is not confirmed. The mass of the object can be inferred from radial-velocity measurements of the star orbited by the body. Perturbations of the radial velocity of a star caused by an exoplanet do not exceed  15 m\,s$^{-1}$ according to the \cite{NASA}. High-resolution spectra are needed to measure such velocities. However, the stars studied are rather faint: \mbox{$m=14^{\rm m}$--$15^{\rm m}$} for \mbox{SOI-3}, \mbox{SOI-5}; \mbox{$m=17^{\rm m}$--$18^{\rm m}$} for \mbox{SOI-4}, SOI-6, and SOI-8, and \mbox{$m\approx19^{\rm m}$} for  SOI-1 and SOI-2~--- their light curves have low $S/N$.

Instead of measuring stellar radial velocities from high-resolution spectra other techniques can be used to test the hypothesis about the
planetary nature, e.g., an analysis of the dependence of the transit depth on photometric band (transmission spectroscopy).
If the transit depths in different filters do not differ then the transiting body can be considered to be gray and the hypothesis that it
may be a star can be rejected. If transit depths differ in different filters  (and does not agree with the
wavelength dependence of limb darkening or the assumption that the exoplanet has atmosphere) then the planetary hypothesis is
not confirmed. Note that one transit event in  SOI-8 was observed with the 50-cm telescope in three filters (Fig.~\ref{fig:SOI_8_post})
however, the number of data points is insufficient to draw conclusions.

Another option is to measure the star's spectrum at different phases with respect to the transit. If the second component is a star of
a significantly different spectral type then the shifts of spectral line should differ at different phases of the orbit of the body.

Yet another option consists in determining the upper limit for the radial velocity of the star from low-resolution
spectra. If the second component is a star then the radial-velocity amplitude of the observed star (several tens of km\,s$^{-1}$)
can be measured, whereas no radial-velocity curve will be observed if the body is a planet.

We plan to use these methods to test the hypothesis about the planetary nature of the exoplanet candidates discovered.

We used modeling  (see Section~4 and  Tab.~\ref{table:params_phys}) to determine for all SOI the radius $R_{{p}} / R_*$ and semi-major axis of the orbit of the second component, $sma / R_*$, in the units of the radius of the star it orbits (Tab.~\ref{table:params_phys}). To determine the size of the second component, we must known the radius of the star studied. According to Vallenari et al. (2022),
the radii of  SOI-1, SOI-2, SOI-4, SOI-6, SOI-7, and \mbox{SOI-8} lie in the \mbox{0.4--0.58} interval; that of SOI-3 is equal to $0.64\,R_{\odot}$ or $1.14\,R_{\odot}$, and the radius of  \mbox{SOI-5} is unknown. Hereafter to test the hypothesis, we use the above estimates of the radii and sizes of orbits of secondary components to a first approximation by formulas~(1) and (2) % (\ref{eq:one}) and (\ref{eq:two})
and compare then with the values of these parameters for confirmed exoplanets.

\begin{table}[H]
	\caption {Radii $R_*$ of the stars according to Gaia DR3 \cite{GAIA2}
		and parameter estimates for exoplanet candidates: the radius  $R_{{p}}$ of the planet and the semi-major axis  $sma$ of its orbit. For SOI-3 the radius from
		Gaia\,EDR\,3 is given in parentheses.
	}
	\centering
	\begin{tabular}{c|c|c|c}
		\hline
		SOI                 & $R_*, R_{\odot}$   & $R_{{p}} / R_*$       & $sma / R_*$    \\
		\hline
		1                   &   0.40            &   0.38            &  7.85     \\
		2                   &   0.55            &   0.42            &  8.47     \\
		3                   &   1.14 (0.64)     &   0.24            &  4.50     \\
		4                   &   0.57            &   0.27            & 13.22     \\
		5                   &   --              &   0.24            & 14.34     \\
		6                   &   0.58            &   0.25            &  9.58     \\
		7                   &   0.52            &   0.39            &  8.24     \\
		8                   &   0.57            &   0.30            &  4.59     \\
		\hline
	\end{tabular}
	\label{table:params_phys}
\end{table}

The radii $R_{{p}}$ of four exoplanet candidates, \mbox{SOI-6}, SOI-4, SOI-1, SOI-8, lie in the $1.39$--$1.64 R_{\mathrm{Jup}}$
interval. The radius of \mbox{SOI-7} candidate is \mbox{$R_{{p}}=1.96\,R_{\mathrm{Jup}}$}, that of \mbox{SOI-3}candidate $R_{{p}}$ is equal to
$1.46\,R_{\mathrm{Jup}}$ or  $2.62\,R_{\mathrm{Jup}}$, and that of exoplanet candidate  SOI-2 is equal to
\mbox{$R_{{p}}=2.22\,R_{\mathrm{Jup}}$}. Currently, there are only two and six exoplanets with the radii $R_{{p}}\!\approx\!2.7\,\!R_{\mathrm{Jup}}$ and
\mbox{$R_{{p}}\!=\!2.0$--$2.2\,\!R_{\mathrm{Jup}}$}, respectively, presented in the \cite{NASA},
with masses estimated to be $M\leq13 M_{\mathrm{Jup}}$,
all other known exoplanets are less massive.
Our inferred estimates of SOI exoplanet candidate radii
lie within the interval of radii of confirmed exoplanets. Thus the eight stars studied can be subdivided into two groups
based on how the radius of the exoplanet candidate is consistent with possible exoplanet radii: SOI-6, \mbox{SOI-4}, SOI-1, SOI-8,
and \mbox{SOI-7} are of greatest interest for further study, followed by SOI-3, SOI-2, and SOI-5.

The minimum orbital period and minimum semi-major axis of known transiting
exoplanets %\footnotemark[\value{footnote}] \linebreak
(with \mbox{$M\leq13\, M_{\mathrm{Jup}}$}) are $P_{\mathrm{min}}=4\,.\!\!^{\rm h}3$  \cite{NASA}
and  \mbox{$sma_{\mathrm{min}}=0.0058$}~AU, respectively. The minimum values of these parameters for the candidates SOI-1--SOI-8 %SOI %SOI-1\dots8
are $P_{\mathrm{SOI8}}=18\,.\!\!^{\rm h}8$ and $sma_{\mathrm{SOI8}}=0.012$~AU, respectively. A total of \linebreak 76~transiting exoplanets
are known to have   \linebreak \mbox{$P_{\mathrm{SOI8}} < P < P_{\mathrm{min}}$}
and 26 (the same except three) have $sma_{\mathrm{SOI8}} < sma < sma_{\mathrm{min}}$.
Hence the periods and orbital semi-major axes of  \mbox{SOI-1--SOI-8} candidates are consistent with the corresponding values for  %known
confirmed exoplanets.\linebreak

Other parameters ($e$, $\Omega$, $i$) determined in the \linebreak process of modeling are of no importance for \linebreak the tested hypothesis.
The orbital eccentricities of \linebreak
\mbox{SOI-1--SOI-7} are greater than 0.3. Less than  3\% of transiting exoplanets are known to have such orbital
eccentricities \cite{NASA}.
The longitude of the ascending node is in no way restricted and it is uniformly distributed in the \mbox{$\Omega=0$--$ 360^\circ$}
interval. These two parameters have the greatest fractional uncertainty. Orbital inclinations $i$ of the candidates discussed
lie in the \mbox{$79\,.\!\!^\circ8$--$87\,.\!\!^\circ6$} interval, which accounts for  26\% of transiting exoplanets \cite{NASA}.

The stellar radii used here (Vallenari et al. 2022)
are inferred from the $B-R$ color index and star parallax assuming that the object is
a single star. Therefore the inferred estimates of the sizes semi-major axes of exoplanets cannot be considered to be sufficiently
reliable. However, at the same time, the available data do not allow us to fully reject the hypothesis about the planetary nature
of the secondary SOI components.

It is also possible to solve the inverse problem of the determination of the maximum stellar radius $R_{\mathrm{*max}}$ given
the $R_{{p}} / R_*$ ratio and adopting  $2.5\,R_{\mathrm{Jup}}$ as the maximum possible radius of an exoplanet.
We thus infer the following maximum radii $R_{\mathrm{*max}}$ for SOI-1--SOI-8:
0.66, 0.61, 1.07, 0.95, 0.66, 1.03, 0.66, and 0.86\,$R_{\odot}$. We can consider rejecting the tested hypothesis
if future radius estimates will exceed the above values.

\section*{Conclusions}
In this study we performed a complete cycle of the search for and analysis of exoplanet candidates by means of transit photometry: photometric monitoring, photometry and searching for transit events in the light curves, predicting upcoming
transits and conducting further photometric observations of candidates, and modeling of transits.

We have discovered eight new exoplanet candidates. The inferred transit parameters, the radii, orbital periods and semi-major axes are consistent with the parameters of the already known exoplanets. We plan to study the spectra of the stars considered and acquire photometric data with the 1-m telescope in order to  more accurately model the transits and determine the parameters of the stars and exoplanet candidates.

The depth of brightness dip during the transit in the stars found corresponds to the cases of giant planets passing across the discs
of small stars \mbox{($0\,.\!\!^{\rm m}06$--$0\,.\!\!^{\rm m}17$)}. This proves that exoplanet observations with the 50-cm robotic telescope
of the Special Astrophysical Observatory of the Russian Academy of Sciences can be used to discover exoplanets larger than Jupiter.
And this is true even if the planetary nature of these exoplanet candidates will not be confirmed in the future.

\section*{Acknowledgments}
This study makes use of the results from the European Space Agency (ESA) space mission Gaia. Gaia data are being processed by the Gaia
Data Processing and Analysis Consortium (DPAC). This study also makes use of the NASA Exoplanet Archive, which is operated by the
California Institute of Technology, under contract with the National Aeronautics and Space Administration under the Exoplanet Exploration
Program.

\section*{Funding}
Authors acknowledge the support of Ministry of Science and Higher Education of the Russian Federation under the grant 075-15-2020-780  \mbox{(No~13.1902.21.0039)}.

\bibliographystyle{apalike}
\bibliography{Eight_can.bib}

\begin{figure*}
	\includegraphics[width=1\textwidth,]{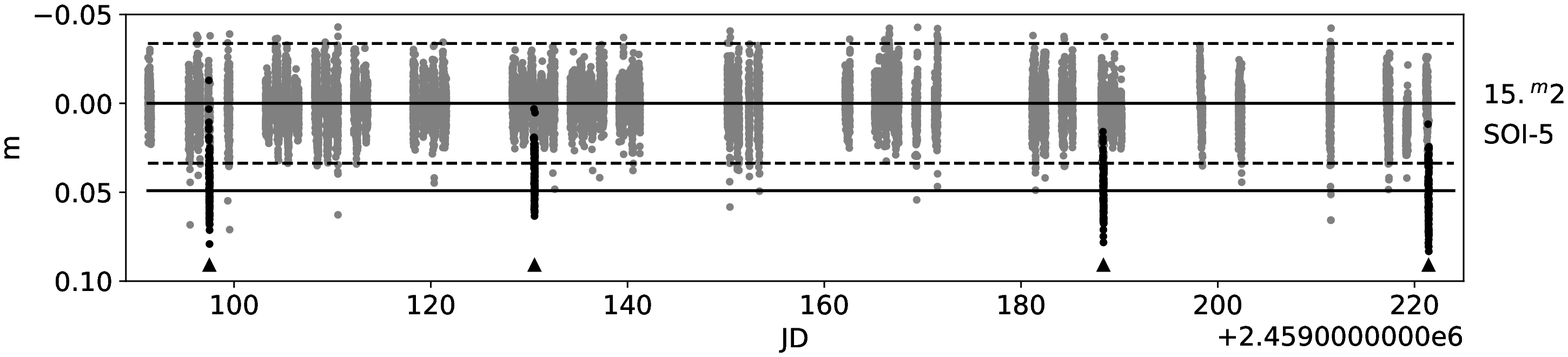}
	\includegraphics[width=1\textwidth,]{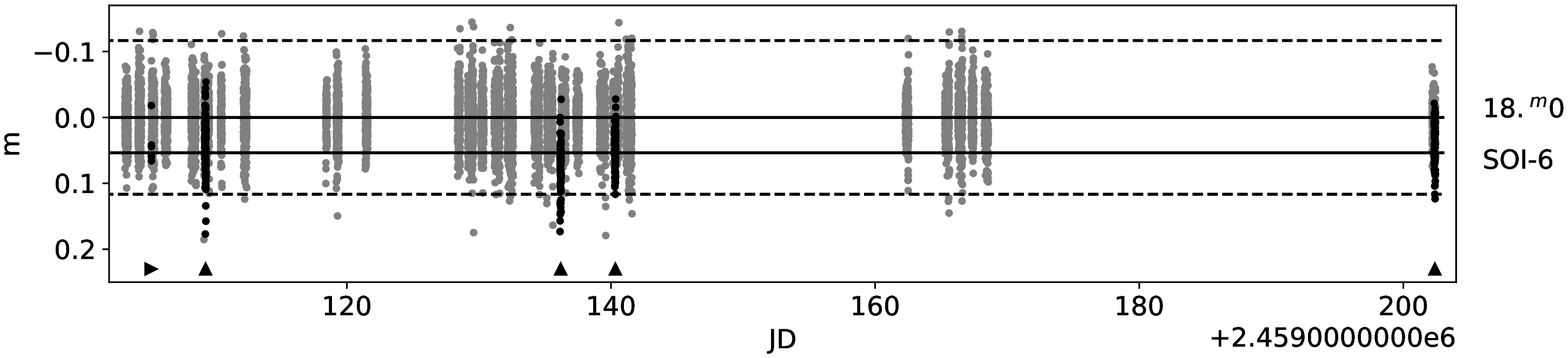}
	\includegraphics[width=1\textwidth,]{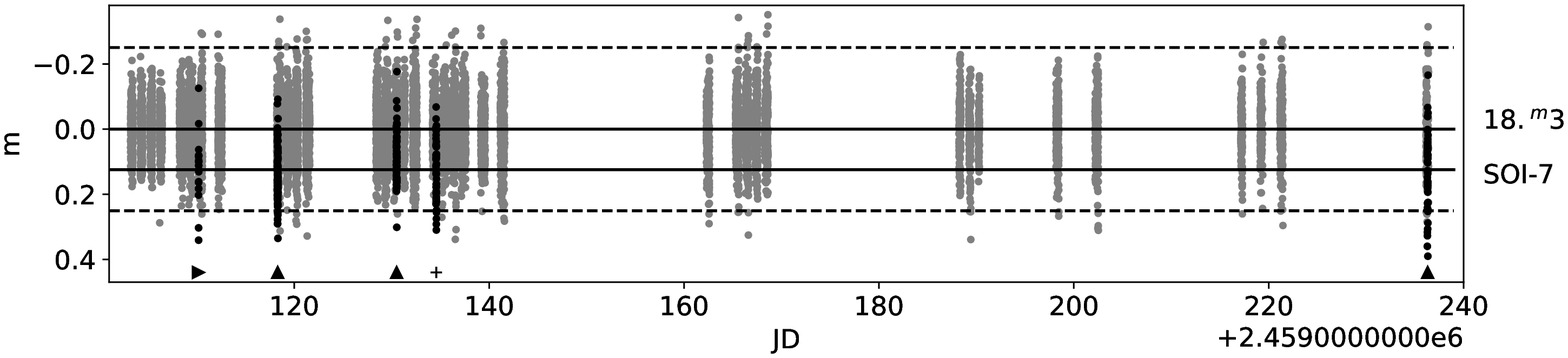}
	\includegraphics[width=1\textwidth,]{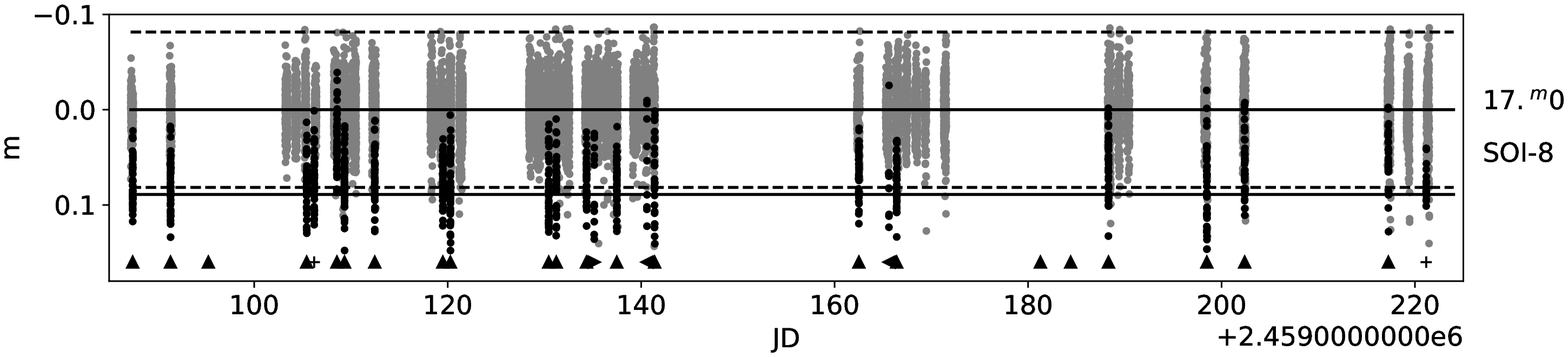}
\end{figure*}

\begin{figure*}
	\includegraphics[scale=0.53]{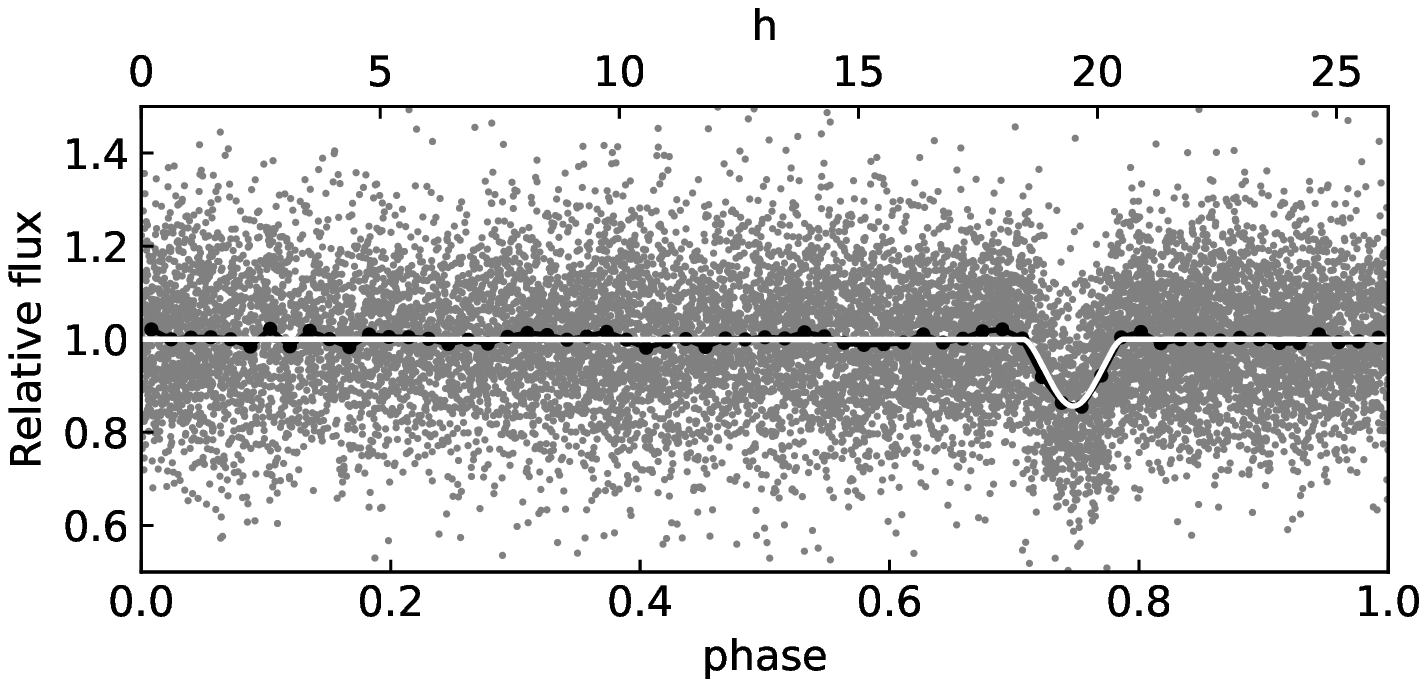}
	\includegraphics[scale=0.4]{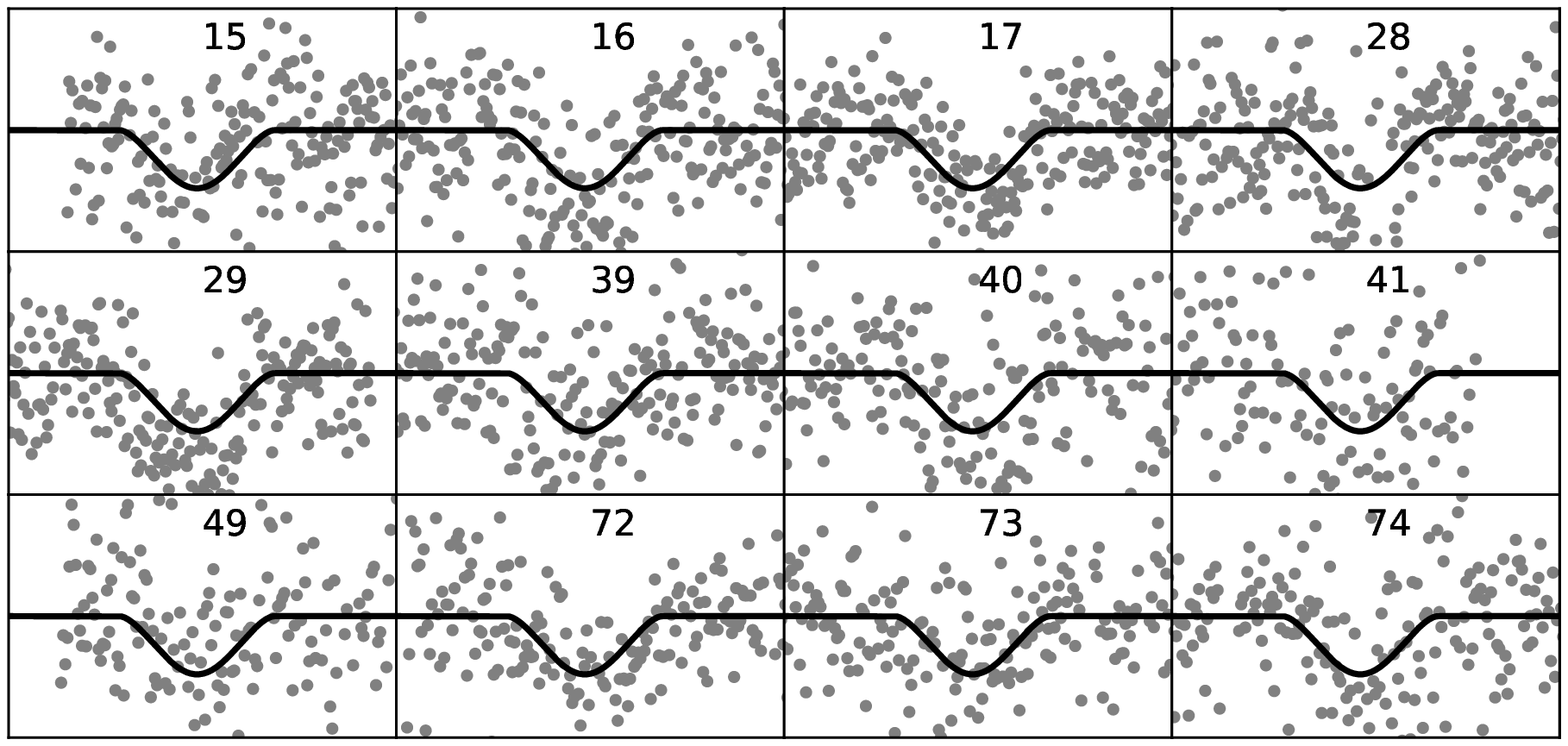}
	\caption{Light curve of SOI-1. Left: phase folded (dots), median values (the black line), and model curve (the white line). Right: fragments for individual transits with the same model curve  (the black line), the period numbers for each transit are indicated. Tick marks and labels are omitted not to clutter the plots.}
	\label{fig:SOI_1}
\end{figure*}

\begin{figure*}
	\includegraphics[scale=0.53, center]{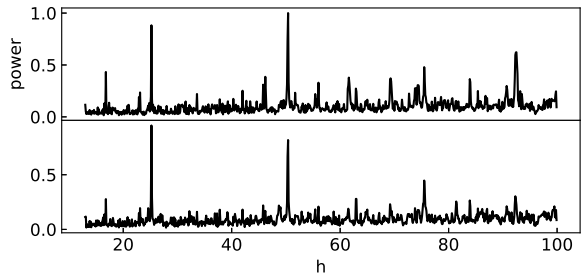}
	\vspace{-5ex} 
	\caption{BLS periodograms for SOI-2 without (bottom) and with (top) subtraction of the nightly trend.}
	\label{fig:SOI_2_PG}
\end{figure*}

\begin{figure*}
	\includegraphics[scale=0.53]{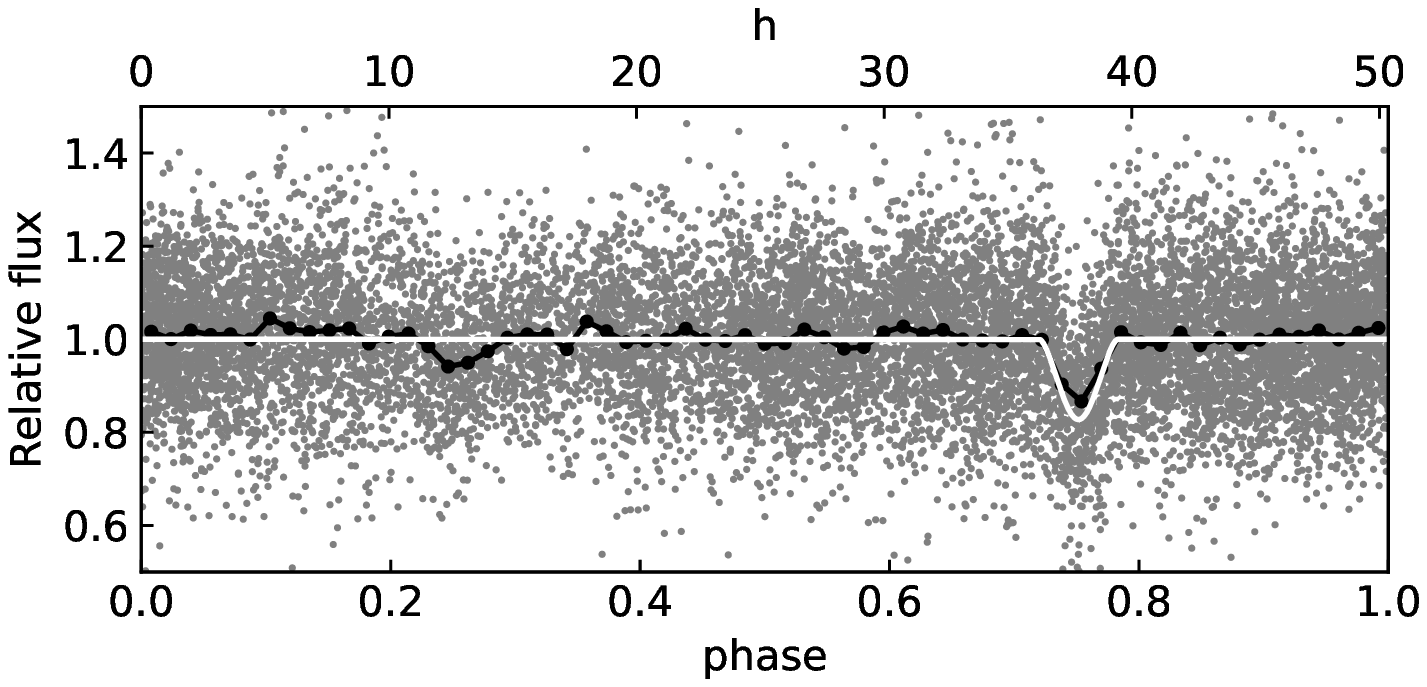}
	\includegraphics[scale=0.5]{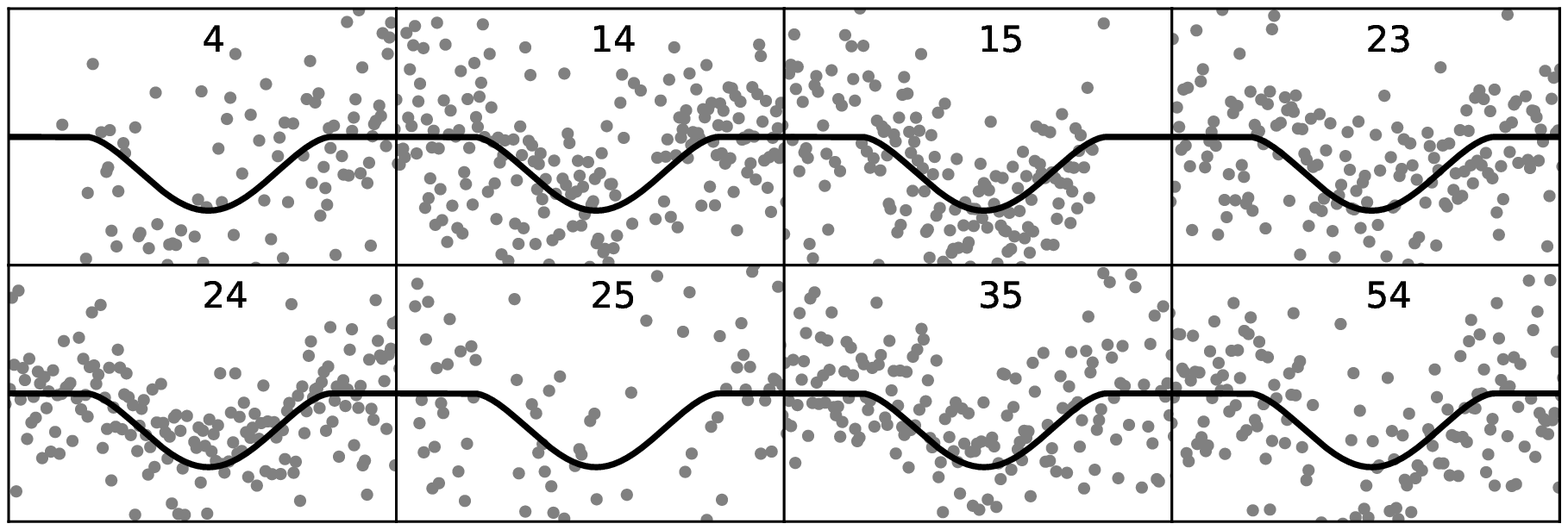}
	\caption{Light curve of SOI-2 (see the caption to Fig.~\ref{fig:SOI_1}).}
	\label{fig:SOI_2}
\end{figure*}

\begin{figure*}
	\includegraphics[scale=0.53]{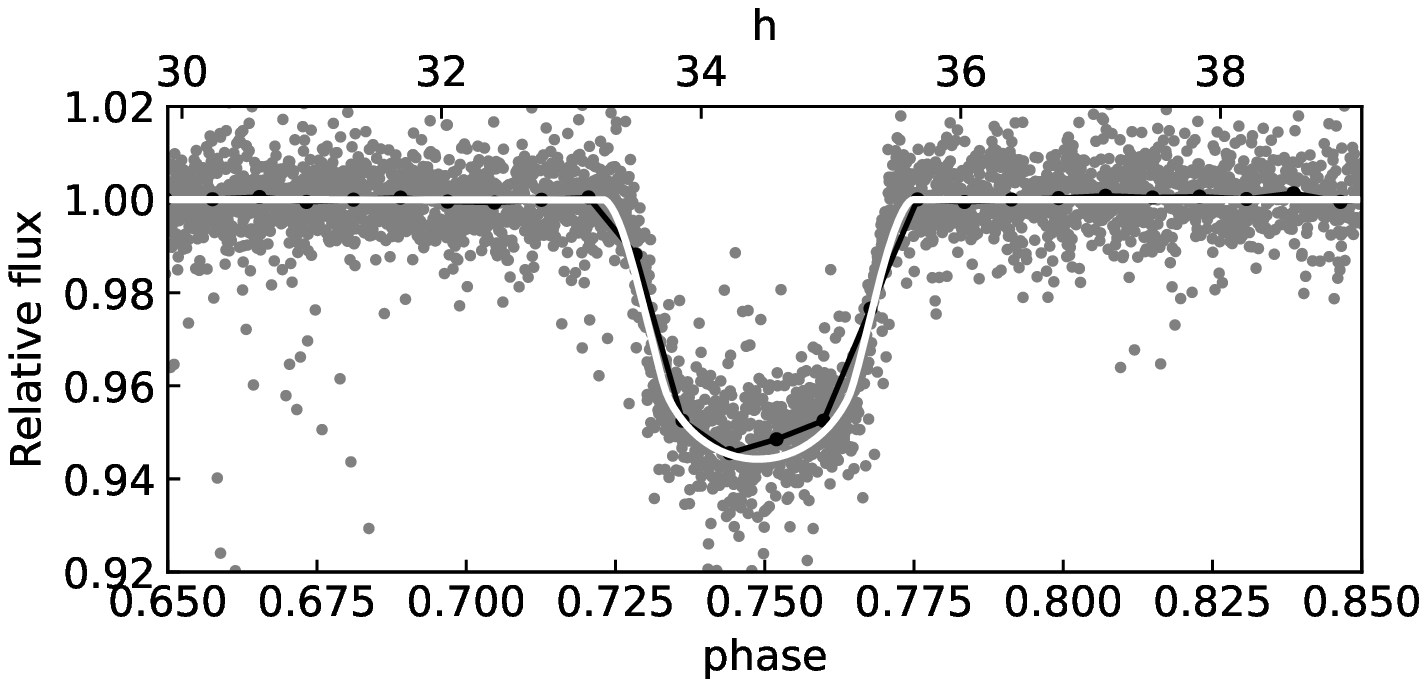}
	\includegraphics[scale=0.4]{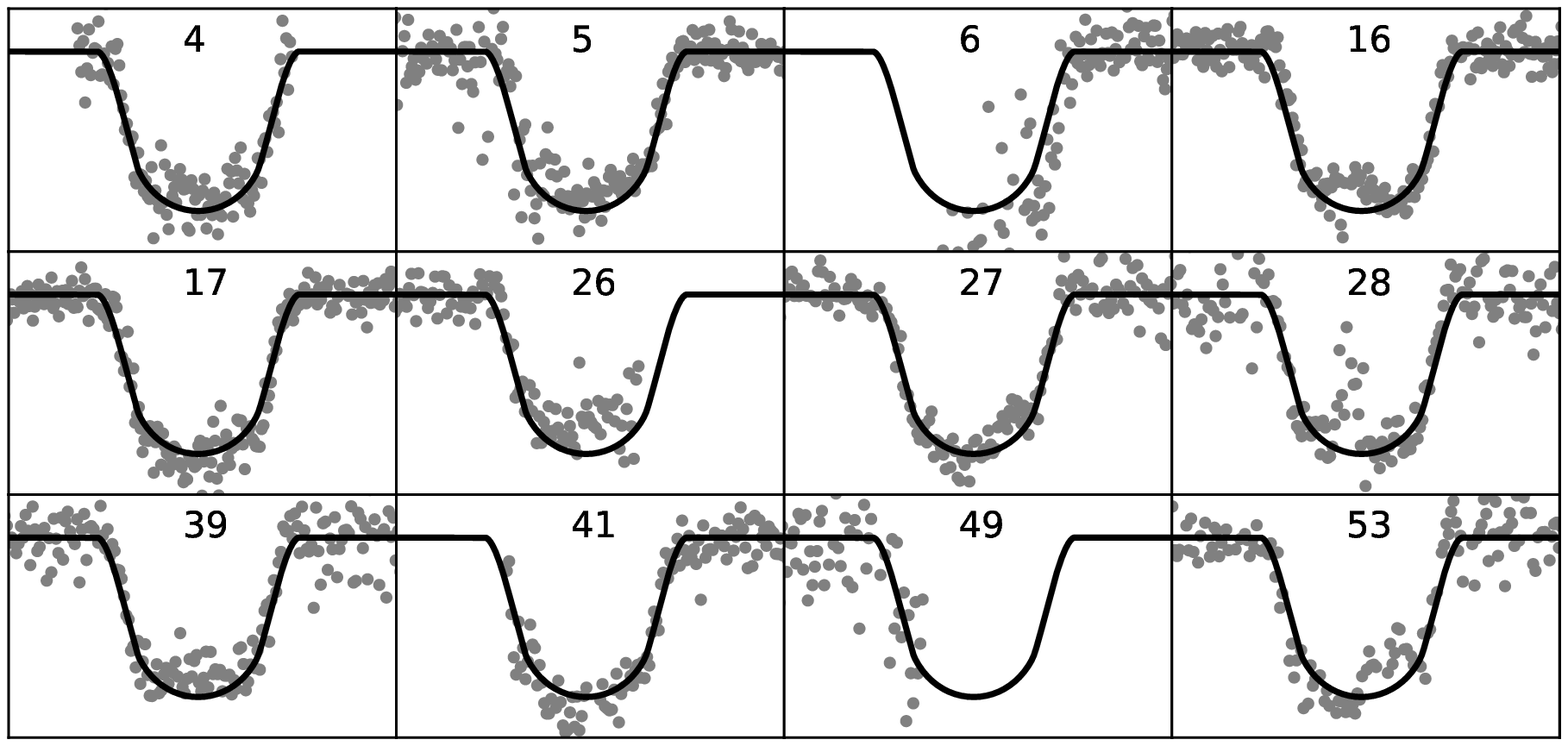}
	\caption{Light curve of SOI-3 (see the caption to Fig.~\ref{fig:SOI_1}).}
	\label{fig:SOI_3}
\end{figure*}

\begin{figure*}
	\includegraphics[scale=0.4, center]{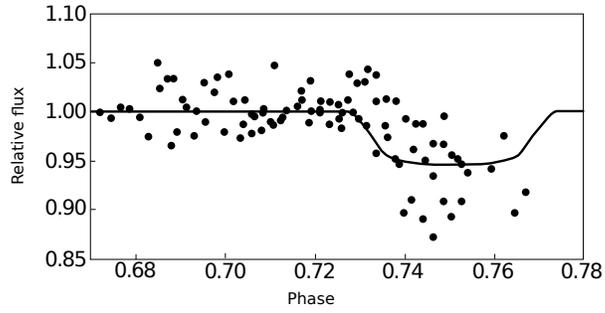}
	\caption{Phased light curve of  SOI-3 based on individual observations made in 2022 ($V$-band filter) and the model curve
		based on the data of 2020 survey-mode observations (see Fig.~\ref{fig:SOI_3}).}
	\label{fig:SOI_3_post}
\end{figure*}

\begin{figure*}
	\includegraphics[scale=0.53]{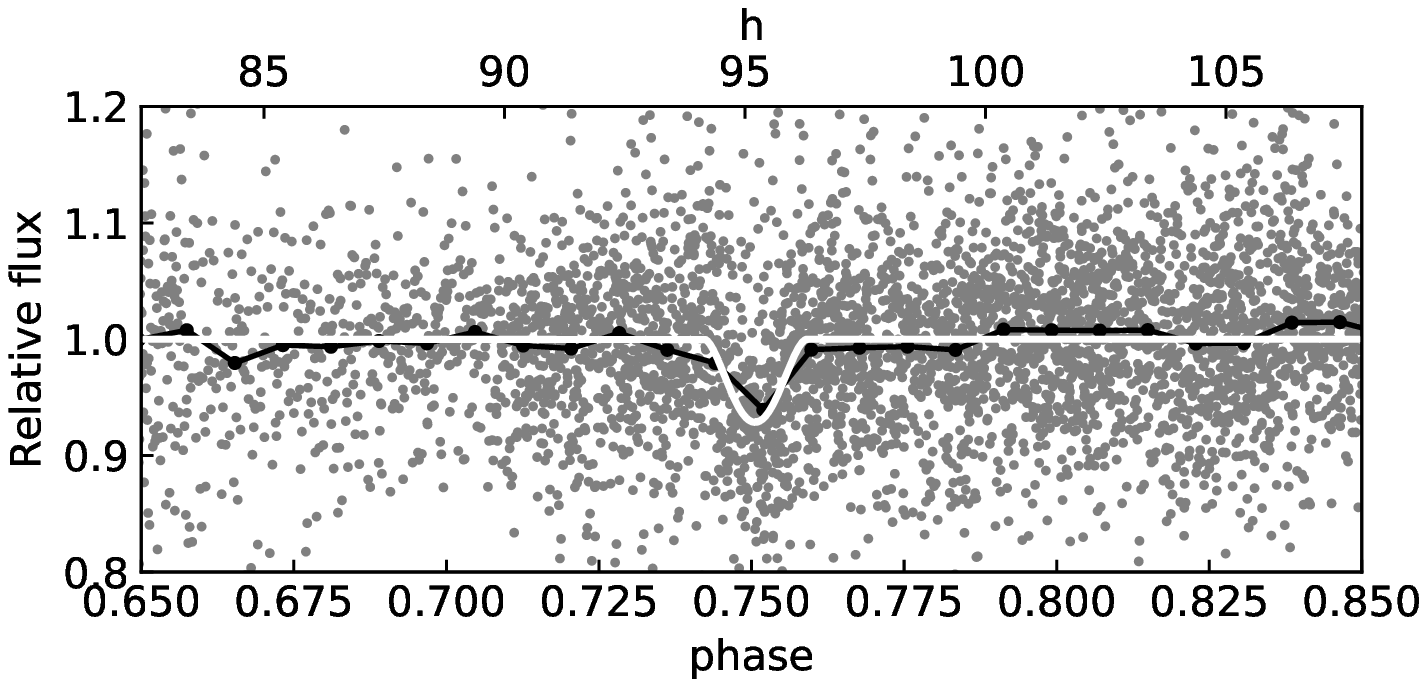}
	\includegraphics[scale=0.5]{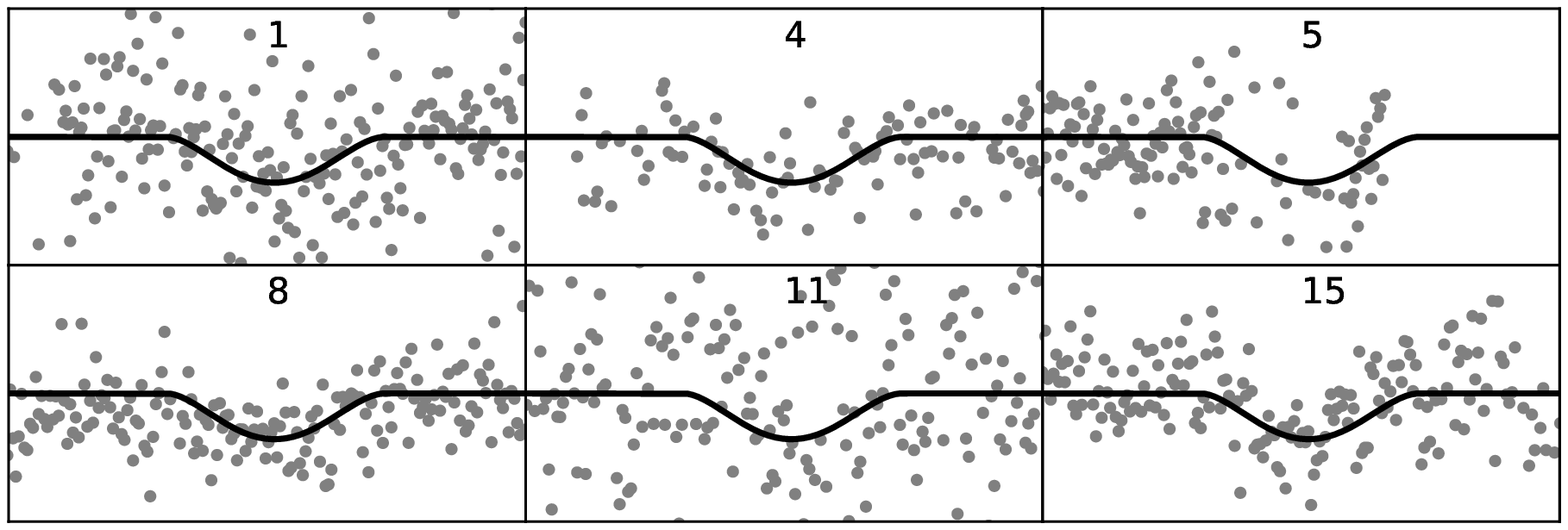}
	\caption{Light curve of SOI-4 (see the caption to Fig.~\ref{fig:SOI_1}).}
	\label{fig:SOI_4}
\end{figure*}
\begin{figure*}
	\includegraphics[scale=0.53]{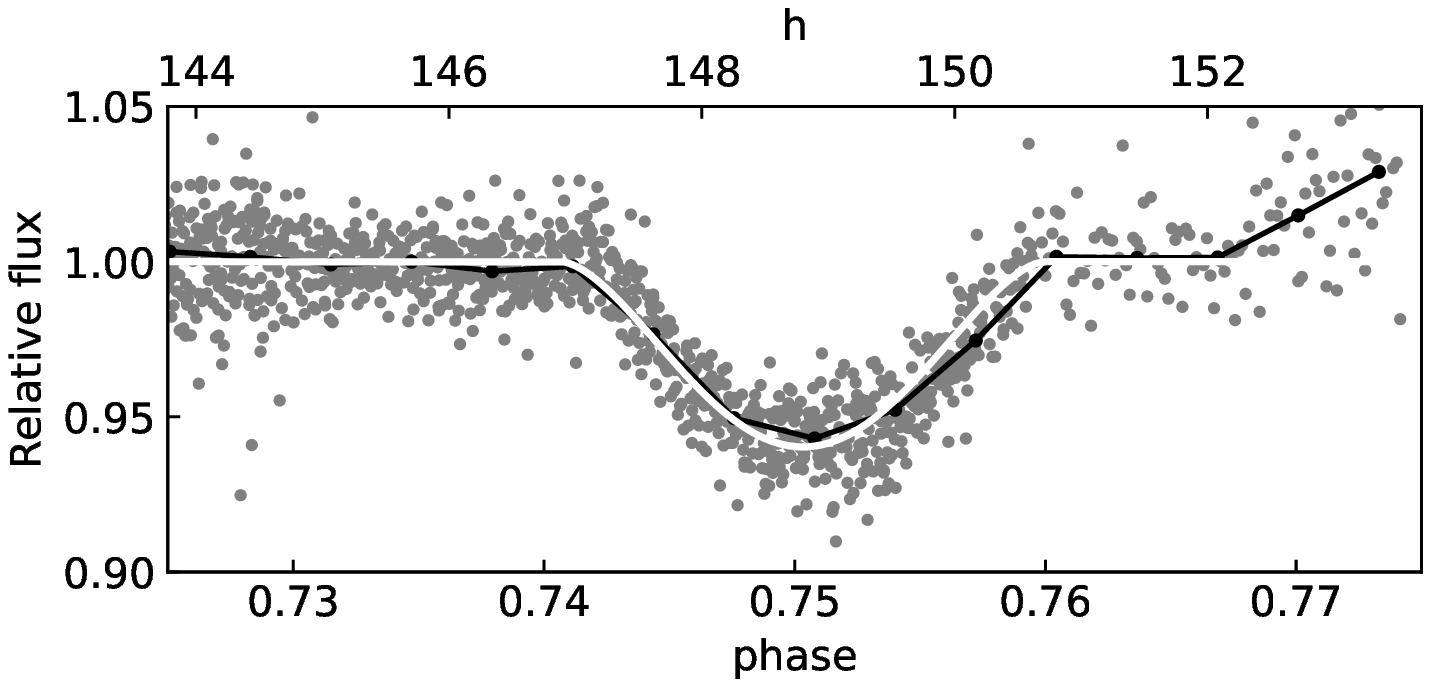}
	\includegraphics[scale=0.5]{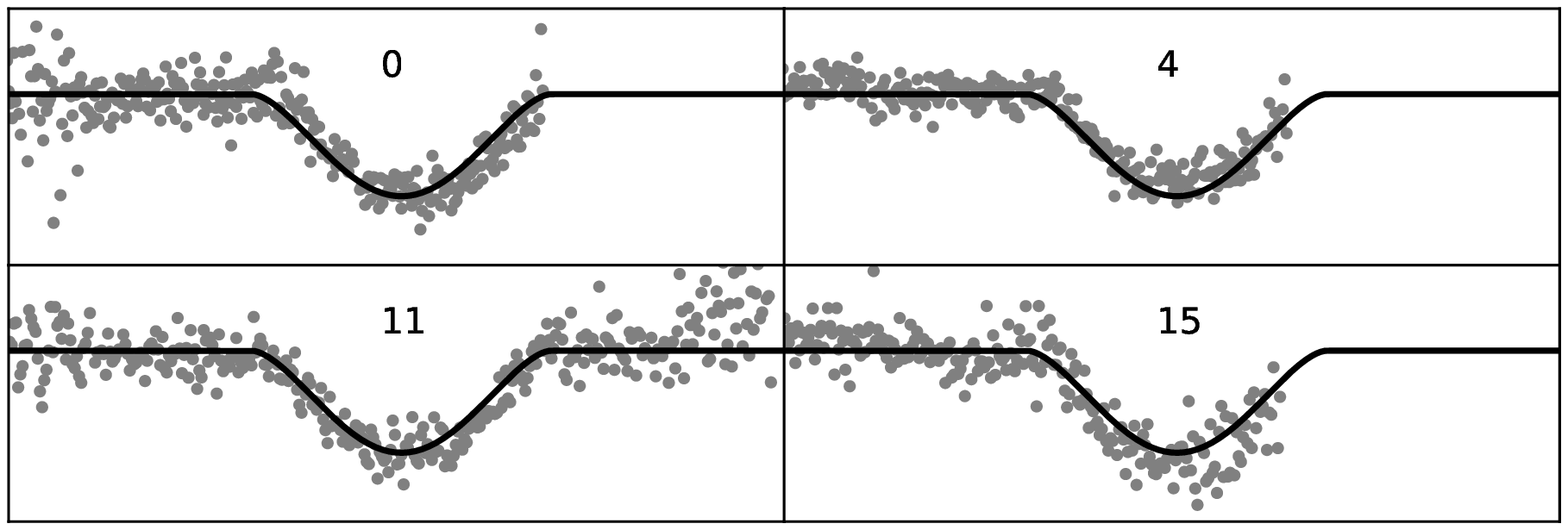}
	\caption{Light curve of SOI-5 (see the caption to Fig.~\ref{fig:SOI_1}).}
	\label{fig:SOI_5}
\end{figure*}
\begin{figure*}
	\includegraphics[scale=0.53]{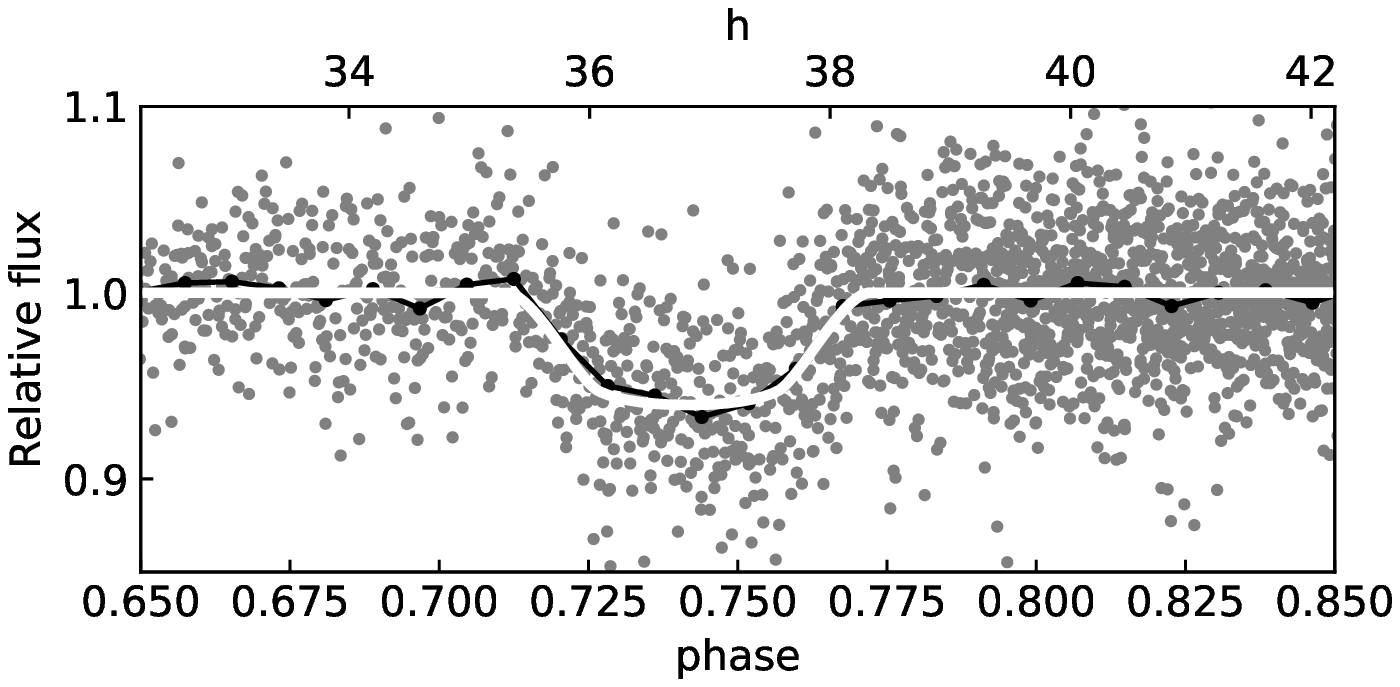}
	\includegraphics[scale=0.5]{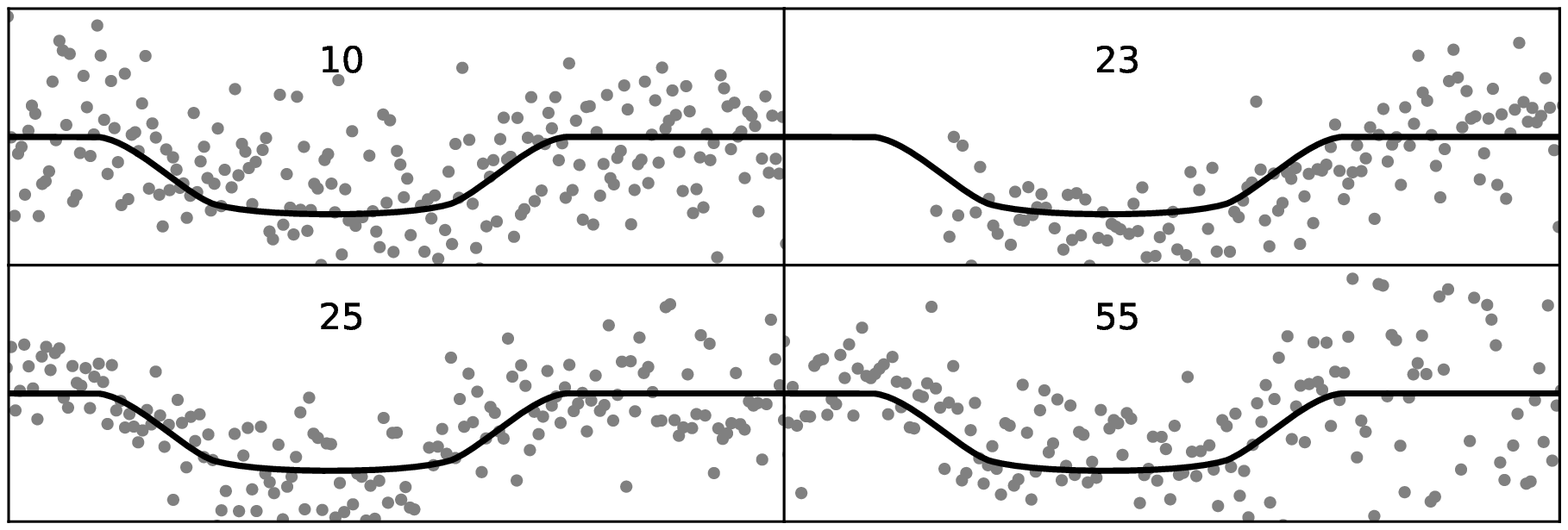}
	\caption{Light curve of SOI-6 (see the caption to Fig.~\ref{fig:SOI_1}).}
	\label{fig:SOI_6}
\end{figure*}

\begin{figure*}
	\includegraphics[scale=0.53]{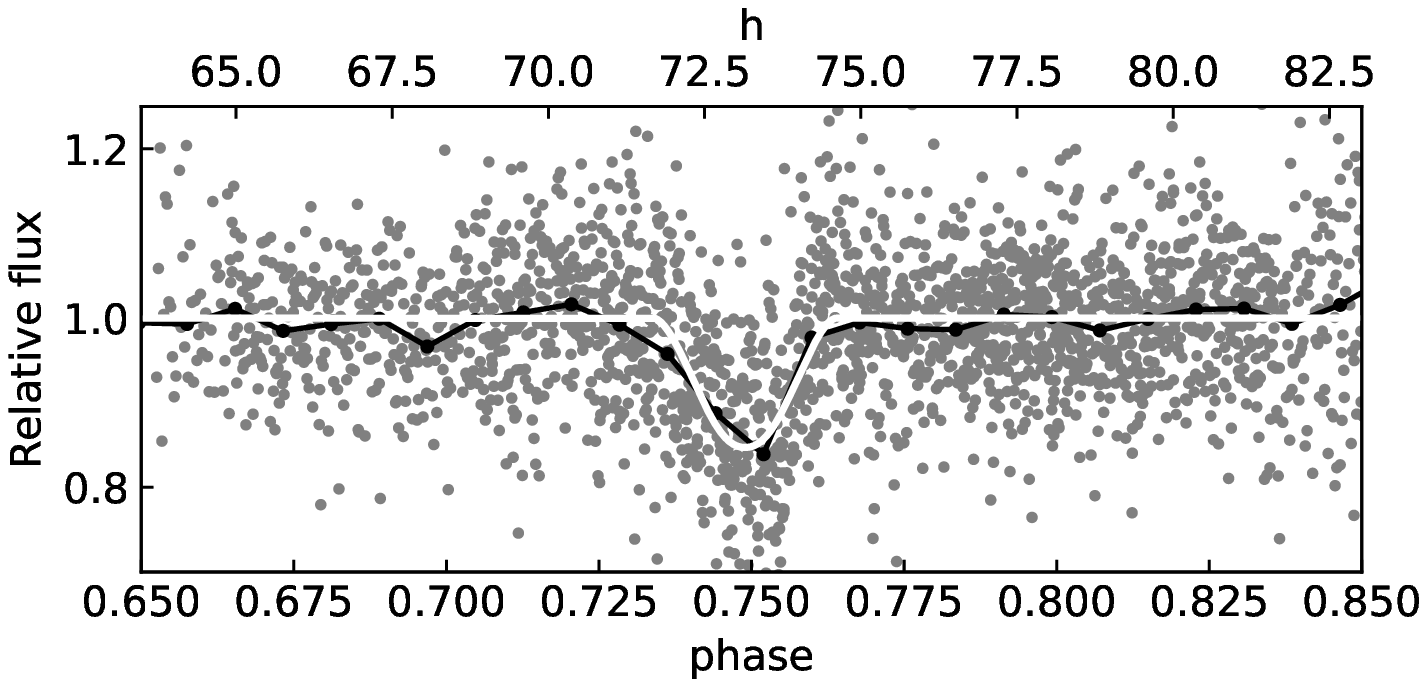}
	\includegraphics[scale=0.5]{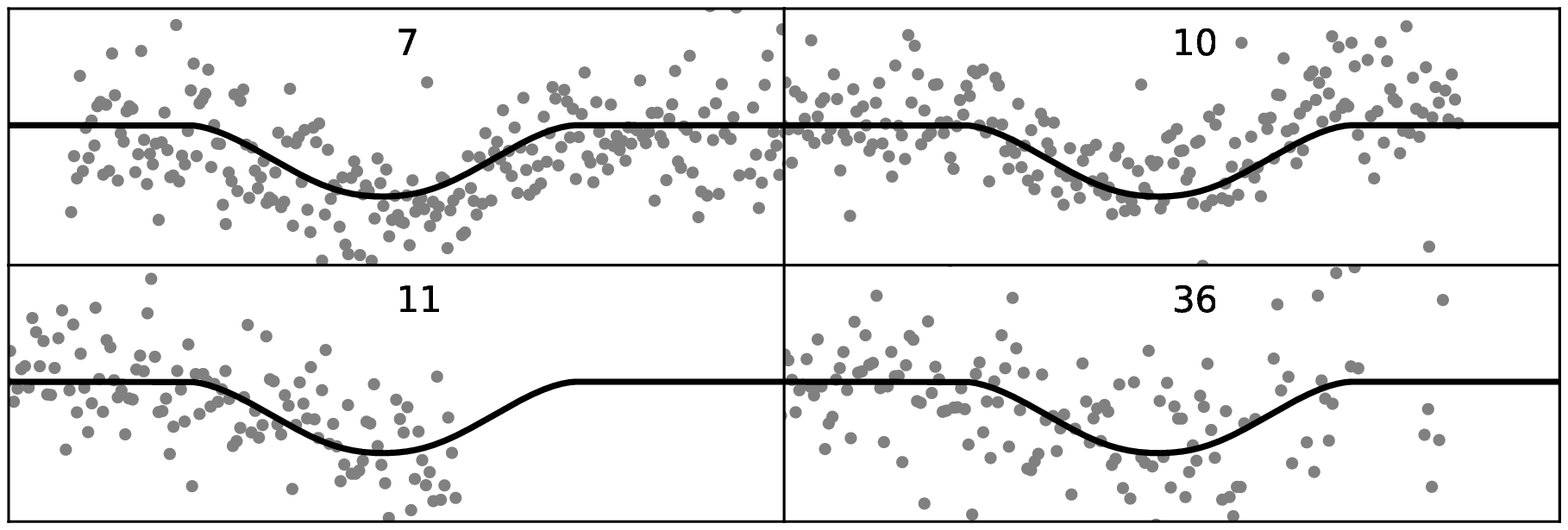}
	\caption{Light curve of SOI-7 (see the caption to Fig.~\ref{fig:SOI_1}).}
	\label{fig:SOI_7}
\end{figure*}

\begin{figure}
	\includegraphics[scale=0.53]{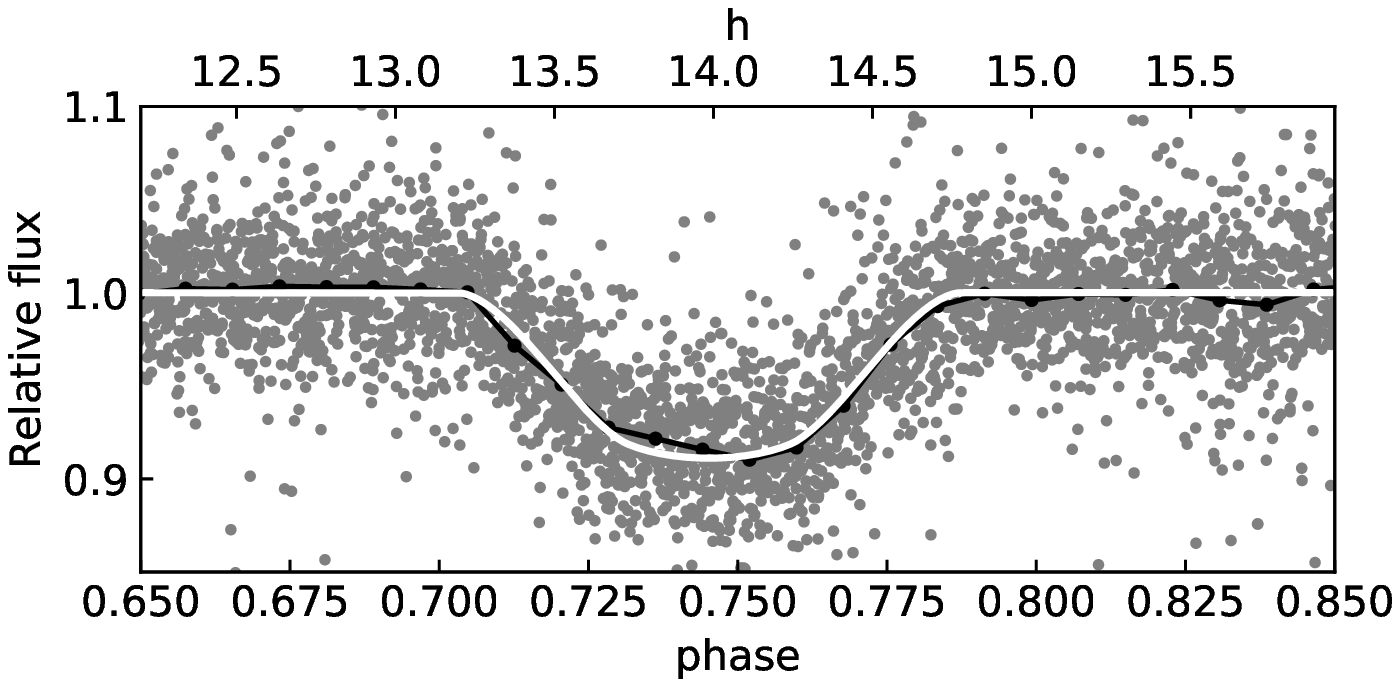}\\
	\includegraphics[scale=0.4]{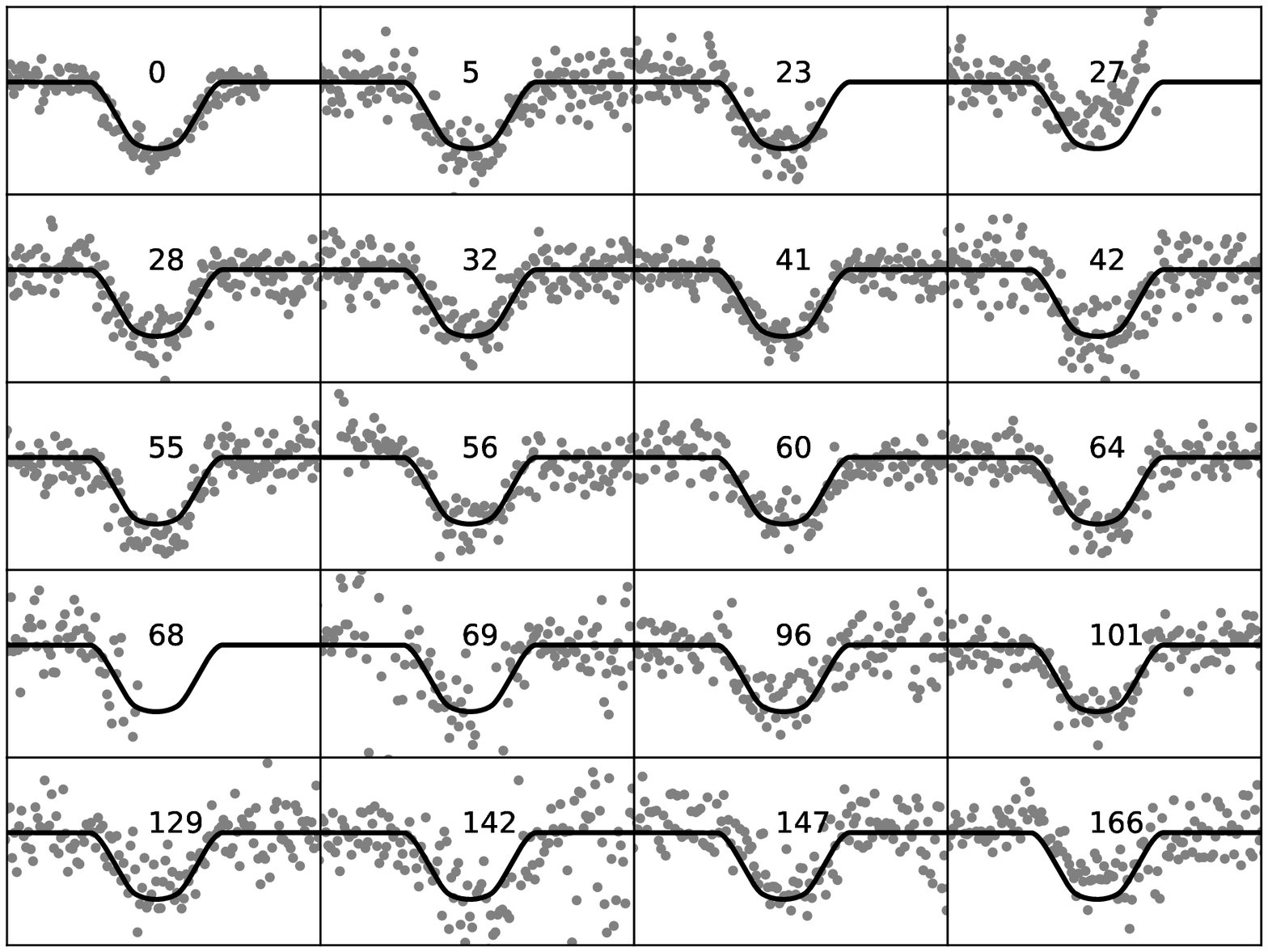}
	\caption{The light curve of  SOI-8 (see the caption to Fig.~\ref{fig:SOI_1}).}
	\label{fig:SOI_8}
\end{figure}

\begin{figure}[H]
	\includegraphics[scale=0.35]{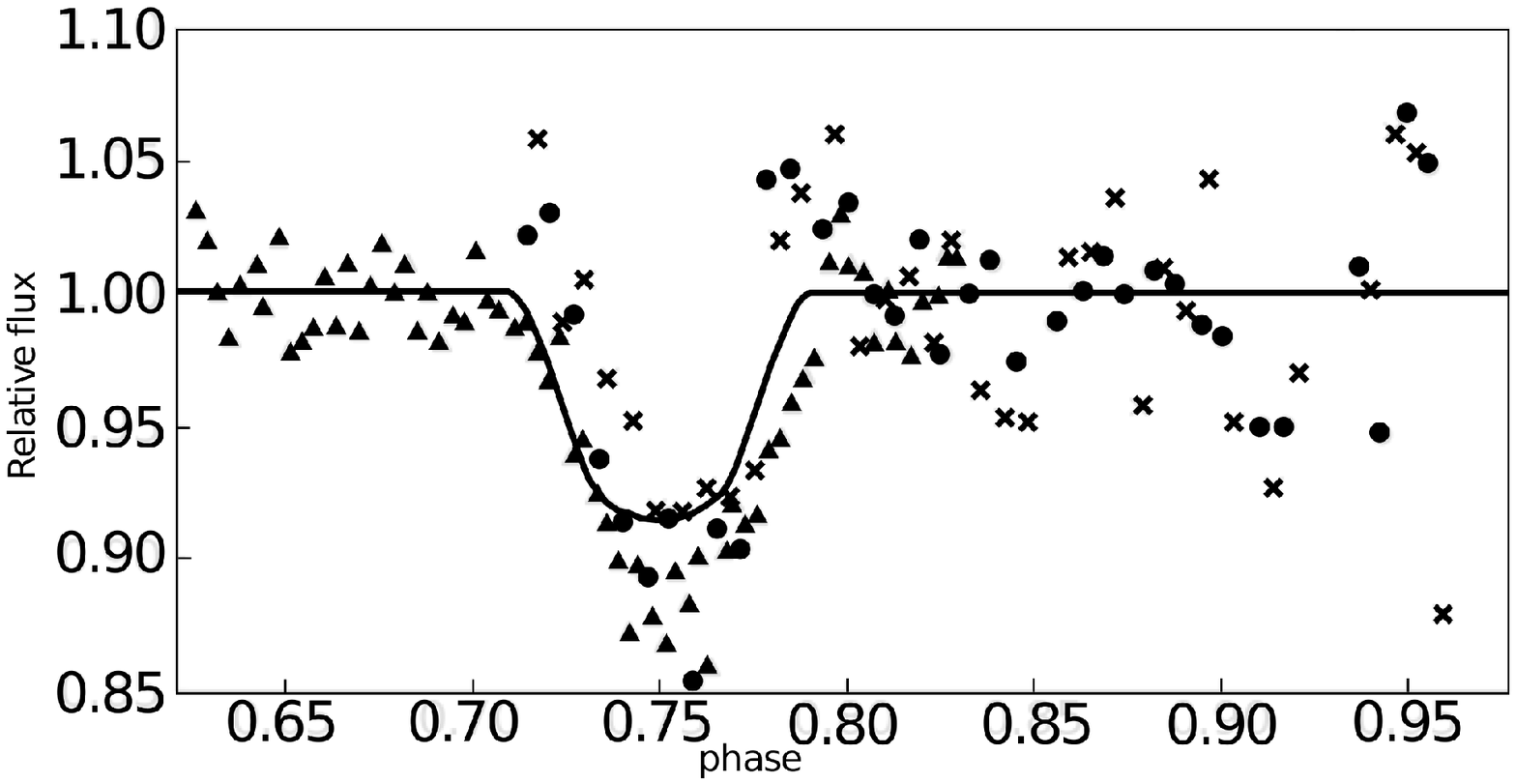}
	\vspace{-2ex} 
	\caption{Phased curve for  SOI-8 based on individual $B$, $V$, $R$ band observations (circles, triangles, and crosses, respectively) in 2022 and the model curve based on the 2020 data (survey-mode observations, see Fig.~\ref{fig:SOI_8}).}
	\label{fig:SOI_8_post}

	\includegraphics[scale=0.45]{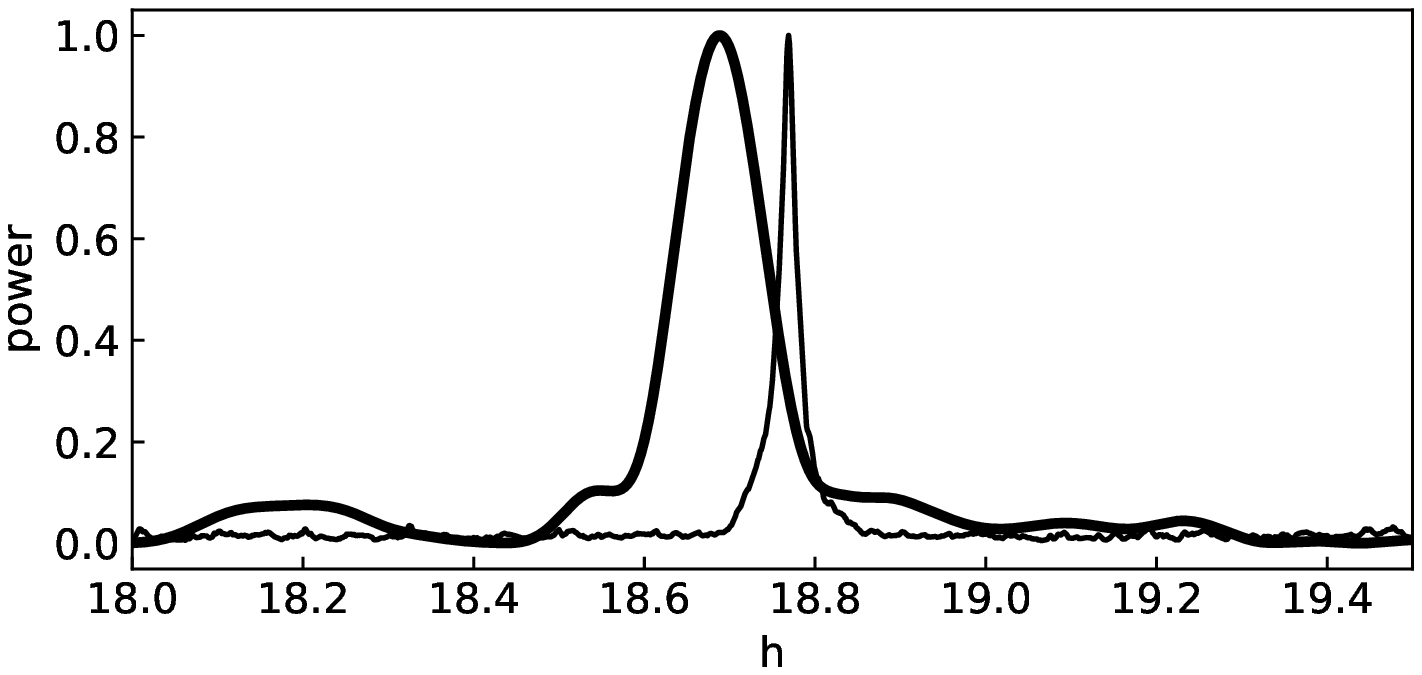}
	\vspace{-2ex} 
	\caption{The BLS periodogram of SOI-8 (the thin line) after the subtraction of the daily trend and the Lomb--Scargle periodogram (the thick line) without the subtraction of this trend and without taking transits into account (data points with phases \mbox{$\phi=0.693$--$0.807$} excluded).
	}
	\label{fig:SOI_8_PG}
	
	\includegraphics[scale=0.5]{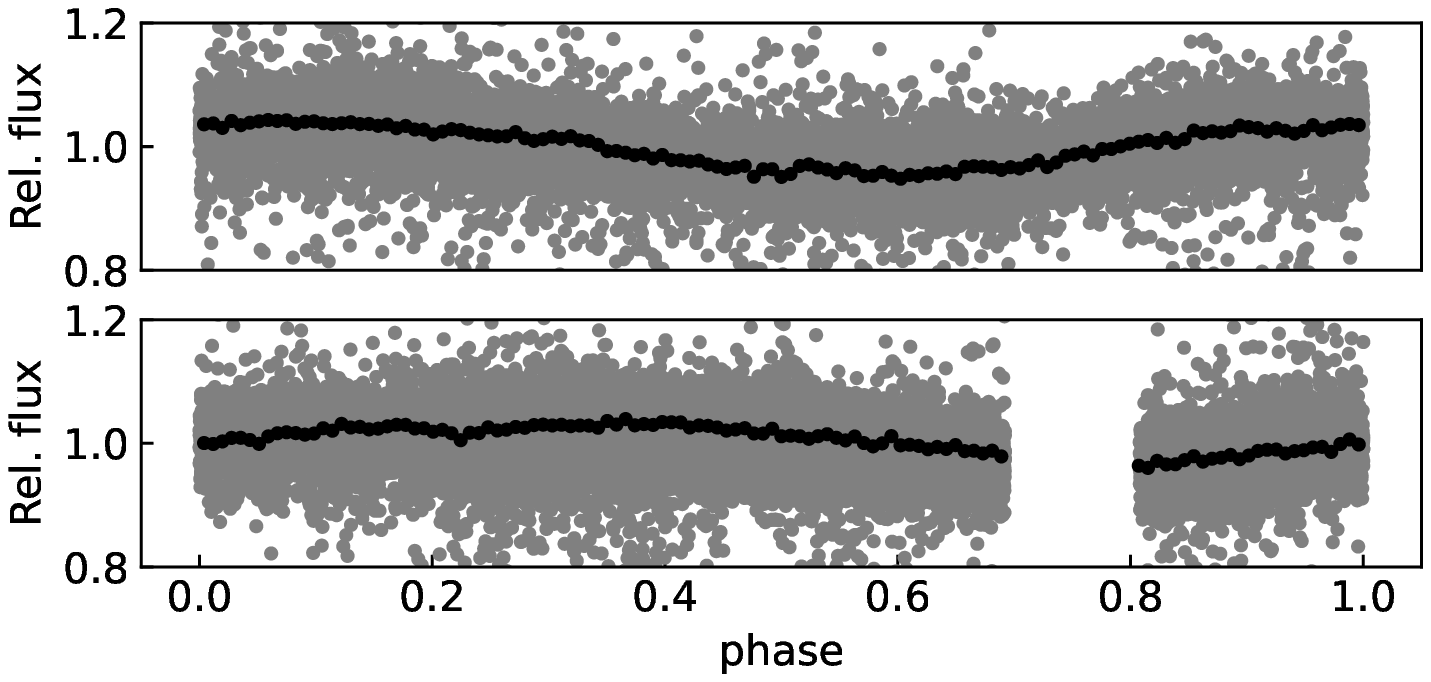}
	\vspace{-2ex} 
	\caption{Phased light curves for  SOI-8 without the subtraction of the trend and without transits
		(data points with phases \mbox{$\phi=0.693$--$0.807$} excluded): folded with the period of $P=18\,.\!\!^{\rm h}688$ (top) and with the	period of transits $P=18\,.\!\!^{\rm h}768$ (bottom).} %
	\label{fig:SOI_8_Var}
\end{figure}

\end{document}